\documentclass{aa}
\usepackage{graphicx}
\usepackage{graphics}
\usepackage[varg]{txfonts}
\usepackage[colorlinks=true,allcolors=blue]{hyperref} 
\usepackage{enumerate}
\usepackage{amsmath}
\usepackage{comment}
\usepackage[para]{threeparttable}
\begin{document}

\title{Old and young stellar populations in DustPedia galaxies and their role in dust heating}

\author{A. Nersesian\inst{1,2,3} \and
E. M. Xilouris\inst{1} \and
S. Bianchi\inst{4} \and
F. Galliano\inst{5} \and 
A. P. Jones\inst{6} \and 
M. Baes\inst{3} \and 
V. Casasola\inst{4,7} \and
L. P. Cassar\`{a}\inst{8}\and
C. J. R. Clark\inst{9} \and 
J. I. Davies\inst{10} \and 
M. Decleir\inst{3} \and
W. Dobbels\inst{3} \and
I. De Looze\inst{3,11}\and
P. De Vis\inst{10} \and
J. Fritz\inst{12} \and
M. Galametz\inst{5} \and 
S. C. Madden\inst{5} \and
A. V. Mosenkov\inst{13,14} \and
A. Tr\v{c}ka\inst{3} \and 
S. Verstocken\inst{3} \and
S. Viaene\inst{3,15} \and
S. Lianou\inst{1,5}}

\institute{National Observatory of Athens, Institute for Astronomy, Astrophysics, Space Applications and Remote Sensing, Ioannou Metaxa and Vasileos Pavlou GR-15236, Athens, Greece \\ \email{a.nersesian@noa.gr} \and  
Department of Astrophysics, Astronomy \& Mechanics, Faculty of Physics, University of Athens, Panepistimiopolis, GR-15784 Zografos, Athens, Greece \and 
Sterrenkundig Observatorium, Universiteit Gent, Krijgslaan 281 S9, 9000 Gent, Belgium \and
INAF - Osservatorio Astrofisico di Arcetri, Largo E. Fermi 5, I-50125, Florence, Italy \and
Laboratoire AIM, CEA/DSM - CNRS - Universit\'{e} Paris Diderot, IRFU/Service d’Astrophysique, CEA Saclay, 91191, Gif-sur- Yvette, France \and
Institut d’Astrophysique Spatiale, UMR 8617, CNRS, Universit\'{e} Paris Sud, Universit\'{e} Paris-Saclay, Universit\'{e} Paris Sud, Orsay, F-91405, France \and
INAF - Istituto di Radioastronomia, Via P. Gobetti 101, 4019, Bologna, Italy \and
INAF - Istituto di Astrofisica Spaziale e Fisica Cosmica, Via Alfonso Corti 12, 20133, Milan, Italy \and
Space Telescope Science Institute, 3700 San Martin Drive, Baltimore, Maryland, 21218, USA \and
School of Physics and Astronomy, Cardiff University, The Parade, Cardiff CF24 3AA, UK \and 
Department of Physics and Astronomy, University College London, Gower Street, London WC1E 6BT, UK \and
Instituto de Radioastronom\'{i}a y Astrof\'{i}sica, UNAM, Campus Morelia, AP 3-72, 58089 Michoac\'{a}n, Mexico \and
Central Astronomical Observatory of RAS, Pulkovskoye Chaussee 65/1, 196140, St. Petersburg, Russia \and 
St. Petersburg State University, Universitetskij Pr. 28, 198504, St. Petersburg, Stary Peterhof, Russia \and 
Centre for Astrophysics Research, University of Hertfordshire, College Lane, Hatfield, AL10 9AB, UK}

\date{Received 23 January 2019 / Accepted 13 March 2019}

\abstract{\textit{Aims}. Within the framework of the DustPedia project we investigate the properties of cosmic dust and its interaction with the stellar radiation (originating from different stellar populations) for 814 galaxies in the nearby Universe, all observed by the \textit{Herschel} Space Observatory. 

\textit{Methods}. We take advantage of the widely used galaxy SED fitting code \textsc{CIGALE}, properly adapted to include the state-of-the-art dust model \textsc{THEMIS}. For comparison purposes an estimation of the dust properties is provided by approximating the emission at far-infrared and sub-millimeter wavelengths with a modified blackbody. Using the DustPedia photometry we determine the physical properties of the galaxies, such as, the dust and stellar mass, the star-formation rate, the bolometric luminosity as well as the unattenuated and the absorbed by dust stellar light, for both the old ($>200~$Myr) and young ($\le200~$Myr) stellar populations.

\textit{Results}. We show how the mass of stars, dust, and atomic gas, as well as the star-formation rate and the dust temperature vary between galaxies of different morphologies and provide recipes to estimate these parameters given their Hubble stage ($T$). We find a mild correlation between the mass fraction of the small a-C(:H) grains with the specific star-formation rate. On average, young stars are very efficient in heating the dust, with absorption fractions reaching as high as $\sim77\%$ of the total, unattenuated luminosity of this population. On the other hand, the maximum absorption fraction of old stars is $\sim24\%$. Dust heating in early-type galaxies is mainly due to old stars, up to a level of $\sim90\%$. Young stars progressively contribute more for `typical' spiral galaxies and they become the dominant source of dust heating for Sm type and irregular galaxies, donating up to $\sim60\%$ of their luminosity to this purpose. Finally, we find a strong correlation of the dust heating fraction by young stars with morphology and the specific star-formation rate.}


\maketitle
%

\section{Introduction} \label{sec:intro}

The spectral energy distribution (SED) of galaxies holds much information on their stellar and dust content. Space observatories such as \textit{Herschel} \citep{2010A&A...518L...1P} and \textit{Spitzer} \citep{2004ApJS..154....1W} provided deep and spatially resolved observations of galaxies in the Local Universe, at infrared (IR) and sub-millimeter (submm) wavelengths, revealing their dust content. Dust has a profound role in shaping the observed SED. In Late-Type Galaxies (LTGs), dust is found to absorb roughly 1/4 to 1/3 of the total energy emitted by stars in the ultraviolet (UV), optical and near infrared (NIR) wavelengths \citep{1991AJ....101..354S, 1995A&A...293L..65X, 2002RvMA...15..239P, 2011ApJ...738...89S, 2016A&A...586A..13V} while redistributing this energy into the mid infrared (MIR), far infrared (FIR) and submm regimes. \citet{2018A&A...620A.112B} investigated the bolometric attenuation by dust, for 814 galaxies of the DustPedia\footnote{\url{http://www.dustpedia.com}} sample \citep{2017PASP..129d4102D, 2018A&A...609A..37C}, as a function of morphological type and luminosity. They found that dust absorbs on average $19\%$ of the total stellar energy budget. However, if only LTGs are considered, the average value increases to $25\%$, which is more in line with previous works in the literature.

The various stellar populations contribute to the heating of the dust grains in a different way. There are two main factors that regulate the level of how efficiently the stellar populations affect the dust heating. One has to do with the spatial distribution of the stars of a specific population with respect to the dust distribution. The general picture for unperturbed Early-Type Galaxies (ETGs) is that all components (young stars, old stars and dust) are distributed in a similar way with a high concentration in the center of the galaxy and with a gradual decrease of the components when going outwards. LTGs, on the other hand, show a central distribution of the old stars in the bulge with the young stars mostly tracing the spiral arms. The second factor is the effective temperature of the stars and their efficiency at heating up the dust in their neighborhood. These factors build up the interstellar radiation field (ISRF) of a galaxy which is responsible for the dust heating. 

An important reason to study the radiation and energy balance of galaxies over the total wavelength range on which they emit, is to learn about their formation and evolution. Physical properties such as the star-formation rate (SFR), the stellar mass ($M_\mathrm{star}$), and the dust mass ($M_\mathrm{dust}$) provide valuable information that can be used to constrain the star-formation history (SFH) and to determine how the baryonic content of galaxies evolved through cosmic time. Many studies have shown that it is possible to approximately estimate these intrinsic physical properties of galaxies by just using single band photometry. Measurements in the NIR wavelengths, for example, being less affected by dust extinction, can be used as proxies for the estimation of the total $M_\mathrm{star}$ of a galaxy \citep{2013MNRAS.433.2946W}. Similarly, measurements in the MIR can trace the SFR adequately \citep{2007ApJ...666..870C, 2015ApJS..219....8C, 2016MNRAS.461..458D}, while FIR emission can provide estimates of $M_\mathrm{dust}$ \citep{2010A&A...518L..51S, 2012MNRAS.419.3505D, 2012MNRAS.425..763G, 2013MNRAS.428.1880A, 2014MNRAS.440..942C}. However, using single band proxies may lead to misleading results if samples of galaxies with mixed morphologies are considered. Accurate determination of these parameters is more complicated and requires a more sophisticated and self-consistent approach. An estimation of galactic properties may be more challenging if objects with entirely different physical parameters show similar SEDs in a given wavelength range. Thus, more advanced methods of SED modelling, that take advantage of the full range of observations from the UV to submm wavelengths are necessary. 

Several codes have been developed in order to model the panchromatic SED of galaxies. Such codes [e.g., \textsc{CIGALE} \citep{2009A&A...507.1793N, 2014ASPC..485..347R, 2019A&A...622A.103B}, \textsc{MAGPHYS} \citep{2008MNRAS.388.1595D}, \textsc{Prospector-$\alpha$} \citep{2017ApJ...837..170L}, \textsc{BEAGLE} \citep{2016MNRAS.462.1415C}, and \textsc{BayeSED} \citep{2014ApJS..215....2H}], take advantage of Bayesian analysis to fit the SEDs of galaxies providing significant information on the actual stellar and dust content, their ability to form stars and the efficiency of the ISRF to heat the dust grains. These codes ensure conservation of energy by allowing re-emission of the stellar light, absorbed by the dust grains, at longer wavelengths. In this work we use the most recent version of \textsc{CIGALE}\footnote{\url{https://cigale.lam.fr}} (version 2018.0) to model the SEDs and to extract the physical properties for the DustPedia galaxies. The code is able to provide both the attenuated and the unattenuated contributions of the different stellar populations (old and young) allowing us to study the fraction of energy that is absorbed by dust for each stellar component. We take advantage of this information to investigate how the stellar populations of galaxies with different morphologies contribute to the dust heating. In addition to \textsc{CIGALE}, we fit single modified black-bodies (MBB) to the FIR-submm observations as an alternative way to estimate the dust temperature and mass.

The data used in this study come from the DustPedia archive\footnote{\url{http://dustpedia.astro.noa.gr}}. The archive was developed within the framework of the DustPedia project (an FP7 funded EU project) providing access to multi-wavelength imagery and photometry for 875 nearby galaxies. In addition to the imagery and the aperture-matched photometry that was applied to all available maps \citep{2018A&A...609A..37C}, additional data such as redshift-independent distances, HI masses \citep{2019A&A...623A...5D}, H$_2$ masses (Casasola et al. 2019, in preparation), optical line and metallicity measurements \citep{2019A&A...623A...5D} as well as information based on 2D photometric fitting such as S\'{e}rsic indices, effective radii, inclination angles, etc. \citep{2019A&A...622A.132M} are also available in the DustPedia archive.

The paper is structured as follows. In Sect.~\ref{sec:sample} we describe the sample of galaxies analyzed in this study. In Sect.~\ref{sec:sed_fitting} we present the SED fitting methods that we use (\textsc{CIGALE} and MBB) along with validation of the quality of the fits. In Sect.~\ref{sec:dsr} we show how the derived main physical parameters vary with galaxy morphology, as parametrized by the Hubble stage ($T$), while in Sect.~\ref{sec:dust_evol} we show the evolution of small a-C(:H) grains with the specific star-formation rate. In Sect.~\ref{sec:stardust} we present how the old and young stellar populations shape the SED of galaxies of different morphological types and their role in dust heating. Our conclusions are summarized in Sect.~\ref{sec:sum_con}. The appendix of this paper is structured as follows. In Appendix~\ref{ap:themis} we present the results of the mock analysis performed by \textsc{CIGALE} using the \textsc{THEMIS} model. In Appendix~\ref{ap:dl14} the results and validation of the fits performed by \textsc{CIGALE} using the widely adopted \citet{2007ApJ...657..810D} model \citep[updated in][hereafter DL14]{2014ApJ...780..172D} are given. In Appendix~\ref{ap:comp}, we compare the main physical galaxy properties derived from the SED modelling with known recipes from the literature. Finally in Appendix~\ref{ap:spline} we give recipes to estimate the main physical properties of galaxies given their Hubble stage.

\section{The Sample} \label{sec:sample}

In our study we make use of multi-wavelength observations of galaxies available in the DustPedia archive \citep{2017PASP..129d4102D, 2018A&A...609A..37C}. The DustPedia sample consists of 875 nearby galaxies (recessional velocities of $<3000$ km s$^{-1}$) with Herschel \citep{2010A&A...518L...1P} detection and angular sizes of $D_{25}>1\arcmin$. The photometry in the DustPedia datasets was carried out in a uniform and consistent way for all observations (up to 41 bands from the FUV to the submm wavelengths) utilizing aperture-matched techniques and robust cross-compatible uncertainty calculations for all bands. For the full description of our photometry pipeline we refer the reader to \citet{2018A&A...609A..37C}. 

The SED fitting routines that we present in the next section require the average flux densities over the filter's relative spectral response function (RSRF) as input. Therefore, we make sure that this requirement is met. The DustPedia photometry conforms to the original pipeline outputs of each instrument/filter combination (or catalog). For UV/optical/NIR data the SED is assumed to be constant over the (relatively narrow) filter bandwidth, hence the flux densities are indeed the average over the filter RSRF. However, for most of the longer wavelength data-points (starting from those of WISE) a spectral shape for the SED is assumed, while for the IRAS bands, a SED with $\nu \times F_\mathrm{\nu}=\text{const.}$. The same convention is used for other bands, with the exception of WISE and MIPS. We first corrected WISE and MIPS data-points from their own color correction to the IRAS constant-energy convention. Then we removed this further colour correction in all long wavelength bands, so that all flux densities give the average over their respective filter RSRF. In most cases, the correction is smaller than a few percent, with the exception of the WISE $12~\mu$m band (whose DustPedia flux densities must be multiplied by 1.05), MIPS bands (an increase by factors 1.03 and 1.07 at 24 and $70~\mu$m, respectively) and IRAS $60~\mu$m (a correction of 0.95).

Before using this photometric dataset we applied several rejection criteria following the guidelines and flagging codes given in \citet{2018A&A...609A..37C} to ensure that only good quality measurements are fed into our modelling. First, we used all bands from GALEX-FUV up to \textit{Planck}-$850~\mu$m, excluding all photometric measurements flagged as contaminated from a nearby galactic or extragalactic source, determined upon visual inspection by \citet{2018A&A...609A..37C}. We furthermore excluded all photometric entries with significant artefacts in the imagery or insufficient sky coverage (leading to a poor background estimate). IRAS and \textit{Planck} data were also checked and measurements where a fraction of the extended emission might have been missed were excluded. Finally, galaxies with insufficient coverage of the SED (i.e. galaxies without fluxes in the wavelength range $0.35 \le \lambda/\mu \text{m} \le 3.6$, and those without fluxes around the peak of the dust emission in the wavelength range $60 \le \lambda/\mu \text{m} \le 500$) were also excluded. With 61 galaxies rejected, our final sample consists of 814 galaxies with the majority ($94\%$) having more than 15 photometric measurements available to constrain the SED modelling performed with \textsc{CIGALE}. 

Galaxies hosting an active galactic nucleus (AGN) require extra treatment with \textsc{CIGALE} \citep[depending on the strength of the AGN emission and on the level that this emission may affect the SED of the galaxy, especially in MIR wavelengths;][]{2015A&A...576A..10C}. Inclusion of AGN templates would significantly increase the required computing time prohibiting us from constructing a dense grid for the rest of the parameters that are significant for the majority of the galaxies in the sample. An assessment on whether or not a galaxy hosts an AGN can be made by using the method described in \citet{2018ApJ...858...38S} and \citet{2018ApJS..234...23A}. The latter method uses a $90\%$-confidence criterion, based on the WISE 3.4 and $4.6~\mu$m bands, to disentangle the galaxies that host an AGN component. \citet{2018A&A...620A.112B}, using this method on the DustPedia galaxies, found that 19 objects, out of the total 814 galaxies, show significant probability in hosting an AGN. These 19 galaxies are: ESO~434-040, IC~0691, IC~3430, NGC~1068, NGC~1320, NGC~1377, NGC~3256, NGC~3516, NGC~4151, NGC~4194, NGC~4355, NGC~5347, NGC~5496, NGC~5506, NGC~7172, NGC~7582, UGC~05692, UGC~06728, and UGC~12690. Since this is only a small fraction ($\sim2\%$) of the DustPedia galaxies we did not use AGN templates in our modelling (in the plots that follow, though, we mark these galaxies with an `X' symbol). Furthermore, we searched for jet-dominated radio galaxies in our sample since synchrotron and free-free emission can be the dominant component in the FIR-submm region of the spectrum. Four such galaxies (NGC~1399, NGC~4261, NGC~4374, and NGC~4486) were found by cross-matching the DustPedia galaxies with the all-sky catalog of radio-galaxies in the local Universe \citep{2012A&A...544A..18V} and marked with an `+' in subsequent plots. In the characteristic case of NGC~4486 (M87), the FIR-submm SED is completely dominated by synchrotron emission \citep{2010A&A...518L..53B}.

Throughout the paper we parameterize the galaxy morphology by the Hubble stage ($T$), the values of which have been retrieved from the HyperLEDA database \citep{2014A&A...570A..13M}\footnote{\url{http://leda.univ-lyon1.fr/}}. A morphological classification in six main sub-classes (E, S0, Sa-Sab, Sb-Sc, Scd-Sdm, and Sm-Irr) is also used wherever specifically indicated.

\section{SED fitting} \label{sec:sed_fitting} 

\begin{table*}[t]
\caption{Parameter grid used for computing the \textsc{CIGALE} templates. A total of 80,041,500 models were produced.}
\begin{center}
\scalebox{0.8}{
\begin{threeparttable}
\begin{tabular}{lc}
\hline 
\hline 
Parameter & Value\\
\hline
Star-Formation History & Flexible Delayed$^\mathrm{(a)}$\\
e-folding time, $\tau_\mathrm{main}$ (Myr) & 500, 750, 1100, 1700, 2600, 3900, 5800, 8800, 13000, 20000\\
galaxy age, $t_\mathrm{gal}$ (Myr) & 2000, 4500, 7000, 9500, 12000\\
quenching or bursting age, $t_\mathrm{flex}$ (Myr) & 200\\
$r_\mathrm{SFR}$ & 0.01, 0.0316, 0.1, 0.316, 1.0, 3.16, 10.0\\
\hline
Stellar population model & BC03$^\mathrm{(b)}$\\
IMF & Salpeter$^\mathrm{(c)}$\\
Metallicity & 0.02\\
\hline
Dust attenuation & Calzetti$^\mathrm{(d)}$\\
Colour excess of the young stars, \textit{E(B-V)} & 0.0, 0.005, 0.0075, 0.011, 0.017, 0.026, 0.038, 0.058, 0.087, 0.13, 0.20, 0.29, 0.44, 0.66, 1.0\\
Reduction factor for \textit{E(B-V)}, \textit{E(B-V)}$_\mathrm{old}$/\textit{E(B-V)}$_\mathrm{young}$ & 0.25, 0.50, 0.75\\
$\delta$  & -0.5, -0.25, 0.0\\
\hline
Dust grain model & \textsc{THEMIS}$^\mathrm{(e)}$; DL14$^\mathrm{(f)}$\\
Fraction of small hydrocarbon solids (\textsc{THEMIS}), $q_\mathrm{hac}$ & 0.02, 0.06, 0.10, 0.14, 0.17, 0.20, 0.24, 0.28, 0.32, 0.36, 0.40\\
PAH abundance [\%] (DL14), $q_\mathrm{PAH}$ & 0.47, 1.12, 1.77, 2.50, 3.19, 3.90, 4.58, 5.26, 5.95, 6.63, 7.32 \\
$U_\mathrm{min}$ & 0.1, 0.15, 0.3, 0.5, 0.8, 1.2, 2.0, 3.5, 6, 10, 17, 30, 50, 80\\
$\alpha$ & 2.0\\
$\gamma$ & 0.0, 0.001, 0.002, 0.004, 0.008, 0.016, 0.031, 0.063, 0.13, 0.25, 0.5\\
\hline \hline
\end{tabular}
\begin{tablenotes}
References: (a) \citet{2016A&A...585A..43C}. (b) \citet{2003MNRAS.344.1000B}. (c) \citet{1955ApJ...121..161S}. (d) \citet{2000ApJ...533..682C}. (e) \citet{2017A&A...602A..46J}. (f) \citet{2014ApJ...780..172D}.
\end{tablenotes}
\end{threeparttable}}
\label{tab:param}
\end{center}
\end{table*}

For the purpose of this work, we make use of the SED fitting code \textsc{CIGALE} to model and interpret the SEDs of the DustPedia galaxies. The code fits the multi-wavelength spectrum of each galaxy in order to derive global properties such as the SFR, the stellar mass $M_\mathrm{star}$, the lower cutoff of the ISRF intensity $U_\mathrm{min}$ and the dust mass $M_\mathrm{dust}^\textsc{CIGALE}$. Furthermore, the stellar component is described by providing the relative contribution of both the young and the old stellar components to the total SED of the galaxy. Complementary to \textsc{CIGALE} we approximate the FIR-submm spectrum of the galaxies with the traditionally used MBB approach with the dust grain properties accordingly scaled to match the \textsc{THEMIS} dust properties. This provides us with an independent estimate of the total dust mass and dust temperature for each galaxy.

\subsection{Fitting the SEDs with \textsc{CIGALE}} \label{subsec:cigale_fit}

\textsc{CIGALE} is an SED modelling tool that allows the user to build galaxy SEDs (in the UV to the submm wavelength range) assuming energy conservation between the energy absorbed by the dust and the energy emitted by stars. The SED reconstruction is made by assuming appropriate stellar population libraries for the emission of different stellar populations \citep[e.g.,][] {2003MNRAS.344.1000B} and a SFH while the dust emission puts constraints on the assumed dust attenuation law \citep{2000ApJ...533..682C} and the grain emission parameters. In addition, nebular line and continuum emission are also included in the UV-NIR wavelength range \citep{2011MNRAS.415.2920I}. 

\subsubsection{Implementing \textsc{THEMIS} to \textsc{CIGALE}}

Various models have been developed for interstellar dust during the last few decades. Two of these models that are widely used are the silicate-graphite-PAH (polycyclic aromatic hydrocarbon) model \citep[]{1990A&A...237..215D, 1992A&A...259..614S, 1997ApJ...475..565D, 2001ApJ...554..778L, 2002ApJ...572..232L, 2001ApJ...551..807D, 2007ApJ...657..810D, 2014ApJ...780..172D} and the silicate-core carbonaceous-mantle model \citep{1990A&A...237..215D, 1990QJRAS..31..567J, 1997A&A...323..566L}. The latter model has recently been updated resulting in the \textsc{THEMIS}\footnote{\url{https://www.ias.u-psud.fr/themis/THEMIS\_model.html}} (The Heterogeneous Evolution Model for Interstellar Solids) dust model \citep{2013A&A...558A..62J, 2017A&A...602A..46J, 2014A&A...565L...9K}. \textsc{THEMIS} was built upon the optical properties of amorphous hydrocarbon and amorphous silicate materials that have been measured in the laboratory \citep{2012A&A...540A...1J, 2012A&A...540A...2J, 2012A&A...542A..98J, 2013A&A...558A..62J, 2017A&A...602A..46J, 2014A&A...565L...9K}. Within this framework, dust is mainly comprised of large carbon-coated amorphous silicate grains and small hydro-carbonaceous grains. The primary goal of the model is to explain the nature of dust in the diffuse interstellar medium (ISM). \textsc{THEMIS} successfully explains the observed FUV-NIR extinction and the shape of the IR to mm dust thermal emission. The model has also been successfully compared to the latest available measures of the diffuse ISM dust extinction and emission in the Milky-Way \citep{2015A&A...577A.110Y, 2015A&A...580A.136F}. Furthermore, it is able to predict the observed relationship between the \textit{E(B-V)} colour excess and the inferred submm opacity derived from \textit{Planck}-HFI observations \citep{2015A&A...577A.110Y, 2015A&A...580A.136F}.

Within the DustPedia framework, we modified \textsc{CIGALE} accordingly to include \textsc{THEMIS} as a separate module for the dust emission parameters. Our approach to generate the template files is similar to the DL14 model \citep{2007ApJ...657..810D, 2014ApJ...780..172D}, i.e., we compute the moments of the average starlight intensity $U$, by using a delta function component and a power-law distribution:

\begin{equation} \label{eq:Uave}
\begin{split}
\left\langle U \right\rangle & = \left(1-\gamma\right)~\int^{U_\mathrm{min}+\Delta U}_{U_\mathrm{min}} U\times\delta\left(U_\mathrm{min}-U\right)~\mathrm{d}U \\
& + \gamma~\int^{U_\mathrm{min}+\Delta U}_{U_\mathrm{min}} U^{1-\alpha}\times \frac{\alpha-1}{\left(U_\mathrm{min}^{1-\alpha} - U_\mathrm{max}^{1-\alpha}\right)} ~\mathrm{d}U, ~\text{for}~\alpha\neq 1 \, ,
\end{split}
\end{equation}

\noindent \citep{2007ApJ...657..810D}, where $\gamma$ is the fraction of the dust heated in photo-dissociation regions (PDR). We created a library of templates based on 3 parameters: (1) the mass fraction of aromatic feature emitting grains, $q_\mathrm{hac}$ (i.e., a-C(:H) smaller than $1.5~$nm), (2) the minimum intensity value of the stellar radiation field that heats the dust, $U_\mathrm{min}$, and (3) the power-law index, $\alpha$. $U_\mathrm{min}$ shares the same parameter space as the DL14 model (0.1-50), with an additional upper limit for \textsc{THEMIS} at $U_\mathrm{min} = 80$ to cover the most extreme cases, whereas $\alpha$ is fixed to 2 in both cases. The maximum cutoff for the starlight intensity distribution, $U_\mathrm{max}$, is fixed at 10$^7$. The small a-C(:H) component has the same effect as the PAH component in the DL14 model. In the diffuse Galactic ISM, $q_\mathrm{PAH} = 7.7\%$ \citep{2011A&A...525A.103C}, and $q_\mathrm{hac} = 17\%$ \citep{2017A&A...602A..46J}. The only difference between a-C(:H) and PAHs is a scaling factor between the two quantities: $q_\mathrm{PAH} \sim q_\mathrm{hac}/2.2$. Finally, in order to retrieve the total dust mass from the SED templates we normalize with $M_\mathrm{dust}/M_\mathrm{H}=7.4\times10^{-3}$ \citep{2017A&A...602A..46J}.

\begin{table*}[t]
\caption{Mean values of various physical properties of the DustPedia galaxies, for different morphological sub-classes. $M_\mathrm{dust}$ and $T_\mathrm{dust}$ are derived by \textsc{CIGALE} and MBB modelling, SFR and $M_\mathrm{star}$ by \textsc{CIGALE}, while the mass of atomic hydrogen ($M_\mathrm{HI}$) is obtained from the literature \citep[see][]{2019A&A...623A...5D}. The number of objects per morphological bin ($N_\mathrm{obj}$) refers to the parameters derived with \textsc{CIGALE}. In total, 814 galaxies were modelled with \textsc{CIGALE}, 678 out of 814 with a single MBB, and 711 have $M_\mathrm{HI}$ measurements.}
\begin{center}
\scalebox{0.85}{
\begin{tabular}{ccc|ccc|cccc}
\hline 
\hline 
$T$ & Type & $N_\mathrm{obj}$ &
$\log\left(\left\langle \text{SFR}\right\rangle\right)$ &
$\log\left(\left\langle M_\mathrm{star}\right\rangle\right)$ & 
$\log\left(\left\langle M_\mathrm{HI}\right\rangle\right)$ &
$\log\left(\left\langle M_\mathrm{dust}^\textsc{CIGALE}\right\rangle\right)$ &
$\log\left(\left\langle M_\mathrm{dust}^\mathrm{MBB}\right\rangle\right)$ &
$\left(\left\langle T_\mathrm{dust}^\textsc{CIGALE}\right\rangle\right)$ &
$\left(\left\langle T_\mathrm{dust}^\mathrm{MBB}\right\rangle\right)$\\
 &  &  & [M$_{\odot}$/yr] & [M$_{\odot}$] & [M$_{\odot}$] & [M$_{\odot}$] & [M$_{\odot}$] & [K] & [K]\\
\hline
  -5 & E       & 51 & -1.27 $\pm$ 0.21 & 10.92 $\pm$ 0.05 & 8.53 $\pm$ 0.01 & 6.15 $\pm$ 0.38 & 6.55 $\pm$ 0.42 & 27.59 $\pm$ 4.91 & 18.62 $\pm$ 5.06\\
  -4 & E$^+$   & 20 & -1.05 $\pm$ 0.14 & 10.94 $\pm$ 0.05 & 9.95 $\pm$ 0.01 & 6.60 $\pm$ 0.27 & 6.66 $\pm$ 0.36 & 26.85 $\pm$ 5.77 & 19.29 $\pm$ 6.95\\
  -3 & S0$^-$  & 34 & -0.59 $\pm$ 0.14 & 10.35 $\pm$ 0.09 & 8.51 $\pm$ 0.04 & 5.95 $\pm$ 0.28 & 6.11 $\pm$ 0.31 & 26.70 $\pm$ 4.23 & 21.87 $\pm$ 5.26\\
  -2 & S0$^0$  & 83 & -0.61 $\pm$ 0.23 & 10.48 $\pm$ 0.07 & 8.86 $\pm$ 0.02 & 6.12 $\pm$ 0.22 & 6.27 $\pm$ 0.30 & 26.89 $\pm$ 4.10 & 21.88 $\pm$ 5.65\\
  -1 & S0$^+$  & 43 & -0.36 $\pm$ 0.14 & 10.46 $\pm$ 0.06 & 8.55 $\pm$ 0.04 & 6.46 $\pm$ 0.13 & 6.56 $\pm$ 0.18 & 24.75 $\pm$ 4.71 & 22.00 $\pm$ 4.76\\
  0  & S0a     & 37 & -0.40 $\pm$ 0.09 & 10.74 $\pm$ 0.04 & 9.24 $\pm$ 0.02 & 6.74 $\pm$ 0.16 & 6.84 $\pm$ 0.16 & 24.63 $\pm$ 4.87 & 20.50 $\pm$ 4.34\\
  1  & Sa      & 50 &  0.03 $\pm$ 0.08 & 10.65 $\pm$ 0.08 & 9.37 $\pm$ 0.02 & 7.00 $\pm$ 0.09 & 7.01 $\pm$ 0.13 & 23.23 $\pm$ 4.34 & 22.89 $\pm$ 4.30\\
  2  & Sab     & 40 &  0.42 $\pm$ 0.11 & 10.48 $\pm$ 0.08 & 9.26 $\pm$ 0.02 & 7.01 $\pm$ 0.07 & 7.02 $\pm$ 0.10 & 22.79 $\pm$ 4.22 & 22.63 $\pm$ 3.65\\
  3  & Sb      & 58 &  0.36 $\pm$ 0.13 & 10.55 $\pm$ 0.08 & 9.45 $\pm$ 0.02 & 7.23 $\pm$ 0.07 & 7.23 $\pm$ 0.11 & 22.62 $\pm$ 3.45 & 22.48 $\pm$ 3.24\\
  4  & Sbc     & 63 &  0.47 $\pm$ 0.10 & 10.40 $\pm$ 0.10 & 9.62 $\pm$ 0.04 & 7.31 $\pm$ 0.06 & 7.33 $\pm$ 0.09 & 21.82 $\pm$ 3.00 & 21.45 $\pm$ 3.82\\
  5  & Sc      & 70 &  0.31 $\pm$ 0.07 & 10.28 $\pm$ 0.09 & 9.66 $\pm$ 0.03 & 7.29 $\pm$ 0.07 & 7.30 $\pm$ 0.11 & 22.09 $\pm$ 2.64 & 21.70 $\pm$ 2.34\\
  6  & Scd     & 84 & -0.05 $\pm$ 0.09 &  9.83 $\pm$ 0.11 & 9.40 $\pm$ 0.02 & 6.94 $\pm$ 0.09 & 6.96 $\pm$ 0.14 & 21.03 $\pm$ 3.30 & 20.21 $\pm$ 2.66\\
  7  & Sd      & 46 & -0.14 $\pm$ 0.05 &  9.62 $\pm$ 0.11 & 9.42 $\pm$ 0.01 & 6.81 $\pm$ 0.11 & 6.84 $\pm$ 0.19 & 20.92 $\pm$ 3.29 & 19.65 $\pm$ 2.77\\
  8  & Sdm     & 32 & -0.56 $\pm$ 0.05 &  9.27 $\pm$ 0.11 & 9.13 $\pm$ 0.02 & 6.46 $\pm$ 0.14 & 6.53 $\pm$ 0.22 & 21.05 $\pm$ 3.94 & 19.17 $\pm$ 3.54\\
  9  & Sm      & 36 & -0.14 $\pm$ 0.04 &  9.24 $\pm$ 0.12 & 9.15 $\pm$ 0.02 & 6.25 $\pm$ 0.17 & 6.42 $\pm$ 0.25 & 24.32 $\pm$ 5.15 & 20.87 $\pm$F 5.51\\
  10 & Irr     & 67 & -0.77 $\pm$ 0.16 &  9.08 $\pm$ 0.14 & 8.88 $\pm$ 0.02 & 5.93 $\pm$ 0.25 & 6.21 $\pm$ 0.44 & 25.11 $\pm$ 5.21 & 19.62 $\pm$ 5.05\\
\hline
$\left[-5.0,-3.5\right)$ & E       & 71  & -1.20 $\pm$ 0.18 & 10.92 $\pm$ 0.05 & 9.36 $\pm$ 0.01 & 6.33 $\pm$ 0.32 & 6.59 $\pm$ 0.40 & 27.38 $\pm$ 5.18 & 19.09 $\pm$ 5.71\\
$\left[-3.5,0.5\right)$  & S0      & 197 & -0.50 $\pm$ 0.16 & 10.52 $\pm$ 0.06 & 8.89 $\pm$ 0.02 & 6.38 $\pm$ 0.17 & 6.51 $\pm$ 0.21 & 25.96 $\pm$ 4.54 & 21.51 $\pm$ 5.16\\
$\left[0.5,2.5\right)$   & Sa-Sab  & 90  &  0.25 $\pm$ 0.10 & 10.58 $\pm$ 0.08 & 9.32 $\pm$ 0.02 & 7.00 $\pm$ 0.08 & 7.01 $\pm$ 0.12 & 23.03 $\pm$ 4.29 & 21.92 $\pm$ 4.03\\
$\left[2.5,5.5\right)$   & Sb-Sc   & 191 &  0.38 $\pm$ 0.10 & 10.42 $\pm$ 0.09 & 9.59 $\pm$ 0.03 & 7.28 $\pm$ 0.07 & 7.29 $\pm$ 0.10 & 22.16 $\pm$ 3.04 & 21.64 $\pm$ 2.83\\
$\left[5.5,8.5\right)$   & Scd-Sdm & 162 & -0.14 $\pm$ 0.07 &  9.70 $\pm$ 0.11 & 9.37 $\pm$ 0.01 & 6.84 $\pm$ 0.10 & 6.87 $\pm$ 0.16 & 21.00 $\pm$ 3.43 & 20.09 $\pm$ 2.91\\
$\left[8.5,10.0\right]$  & Sm-Irr  & 103 & -0.44 $\pm$ 0.07 &  9.14 $\pm$ 0.13 & 9.00 $\pm$ 0.02 & 6.07 $\pm$ 0.21 & 6.30 $\pm$ 0.35 & 24.83 $\pm$ 5.21 & 20.55 $\pm$ 5.27\\
\hline \hline
\end{tabular}}
\label{tab:phys_param}
\end{center}
\end{table*}

\subsubsection{The parameter space} \label{subsubsec:par_space}

To model the SED of a galaxy with \textsc{CIGALE}, a parametric SFH has to be assumed. Several types have been proposed (and described with simple analytic functions) to account for the evolution of the stellar content of the galaxies with time. In \citet{2015A&A...576A..10C} three SFHs [an exponentially decreasing SFH ($1\tau-$dec) of e-folding time $\tau$, a combination of two exponentially decreasing SFHs ($2\tau-$dec) and a delayed SFH] are discussed and compared with simulated SFHs from \textsc{GALFORM SAM} \citep{2000MNRAS.319..168C} concluding that the delayed SFH is the better choice for estimating SFR, $M_\mathrm{star}$ and age of galaxies. Recently, \citet{2016A&A...585A..43C, 2017A&A...608A..41C} suggested that a flexible-delayed SFH is able to describe well both field and cluster spirals as well as galaxies with recent starburst activity or SFR decline. Four parameters are used to describe the specific type of SFH, namely, the e-folding time of the main stellar population model ($\tau_\mathrm{main}$), the age of the oldest stars in the galaxy ($t_\mathrm{gal}$), the age of the significant drop or rise in the star-formation activity ($t_\mathrm{flex}$), and the ratio $r_\mathrm{SFR}$ of the SFR after quenching or bursting over the SFR at the time of $t_\mathrm{flex}$ of the star-formation. The SFR as a function of time is then given by:

\begin{equation} \label{eq:sfr}
\text{SFR}(t) \ \propto
\begin{cases}
  t \times \exp(-t/\tau_\mathrm{main}) & ,\text{for} \ t \leq t_\mathrm{flex} \\
  r_\mathrm{SFR} \times \text{SFR}(t=t_\mathrm{flex}) & ,\text{for} \ t > t_\mathrm{flex} \, ,
\end{cases}
\end{equation}

\noindent \citep{2016A&A...585A..43C, 2017A&A...608A..41C}. The SFH implemented in \textsc{CIGALE} is the combination of two stellar components resembling an old and a young stellar population. The two populations roughly represent a burst or quench of star-formation in addition to a more passively evolving stellar component. The old stellar component is modelled with an exponentially decreasing SFR with various values of the e-folding rate $\tau_\mathrm{main}$ and age $t_\mathrm{gal}$ (see Table~\ref{tab:param}). The young stellar component consists of a burst or decline of constant star-formation starting at time $t_\mathrm{flex}$ (in our case $200~$Myr ago) whose amplitude is adjustable and scaled with the ratio $r_\mathrm{SFR}$. In what follows, SFR refers to the current star-formation rate (see Equation~\ref{eq:sfr}). The stellar and nebular emissions are then attenuated using a power–law–modified starburst curve \citep{2000ApJ...533..682C}, extended with the \citet{2002ApJS..140..303L} curve:

\begin{equation} \label{eq:att}
A(\lambda)=\left(A(\lambda)_\mathrm{SB}\times(\lambda/550~\mathrm{nm})^\delta +D_\lambda\right)\times\frac{E\left(B-V\right)_{\delta = 0}}{E\left(B-V\right)_{\delta}} \, ,  
\end{equation} 

\noindent \citep{2009A&A...507.1793N, 2019A&A...622A.103B}. $D_\lambda$ is a Lorentzian–like Drude profile modelling the UV bump at 217.5~nm in the attenuation curve (in our case $D_\lambda = 0$, i.e., no bump), with $\delta$ being a free parameter modifying the slope of the attenuation curve and the last term in Equation~\ref{eq:att} being an attenuation reduction factor for the old stellar population (older than 10~Myr). We chose to use an attenuation law without a bump feature for three reasons: (1) The UV emission in our galaxy sample is covered by just the two GALEX bands. To get a better constrain on the UV bump would require additional data and preferably NUV spectra \citep{2012A&A...545A.141B} which are not available for our sample, (2) we followed the works of \citet{2009A&A...507.1793N}, \citet{2012A&A...539A.145B, 2016A&A...591A...6B} which they also use a bump-less attenuation law, and (3) we did not set the UV bump as an additional free parameter since this would significantly increase computational time. The stellar spectrum is calculated by convolving the \citet{2003MNRAS.344.1000B} single stellar populations with the SFH, assuming a \citet{1955ApJ...121..161S} initial mass function (IMF) and a metallicity of $Z=0.02$, which corresponds to $12+\log(\mathrm{O}/\mathrm{H})=8.86$ assuming the solar metallicity ($Z=0.0134$) and oxygen abundance ($12+\log(\mathrm{O}/\mathrm{H})=8.69$) from \citet{2009ARA&A..47..481A}.

\textsc{CIGALE} uses a Bayesian analysis to derive the galaxy properties. The modelled SEDs are integrated into a set of filters and compared directly to the observations. The observations are assigned with an extra $10\%$ uncertainty (added quadratically to the measured uncertainty) to allow for unknown systematic errors in the object's photometry and the models \citep[see][]{2009A&A...507.1793N}. For each parameter, a probability distribution function (PDF) analysis is carried out. Given the observed SED of a galaxy, \textsc{CIGALE} derives the posterior probability distribution of the physical parameters. The posterior probability is simply the dot product of the prior, i.e., the probability of a model being used before fitting the data, and the likelihood that the data match the model created by the parameter grid. The result of the analysis is the likelihood-weighted mean value of the PDF while the associated error is the likelihood-weighted standard deviation \citep{2019A&A...622A.103B}. 

\begin{figure}[ht]
\centering
\includegraphics[width=9cm]{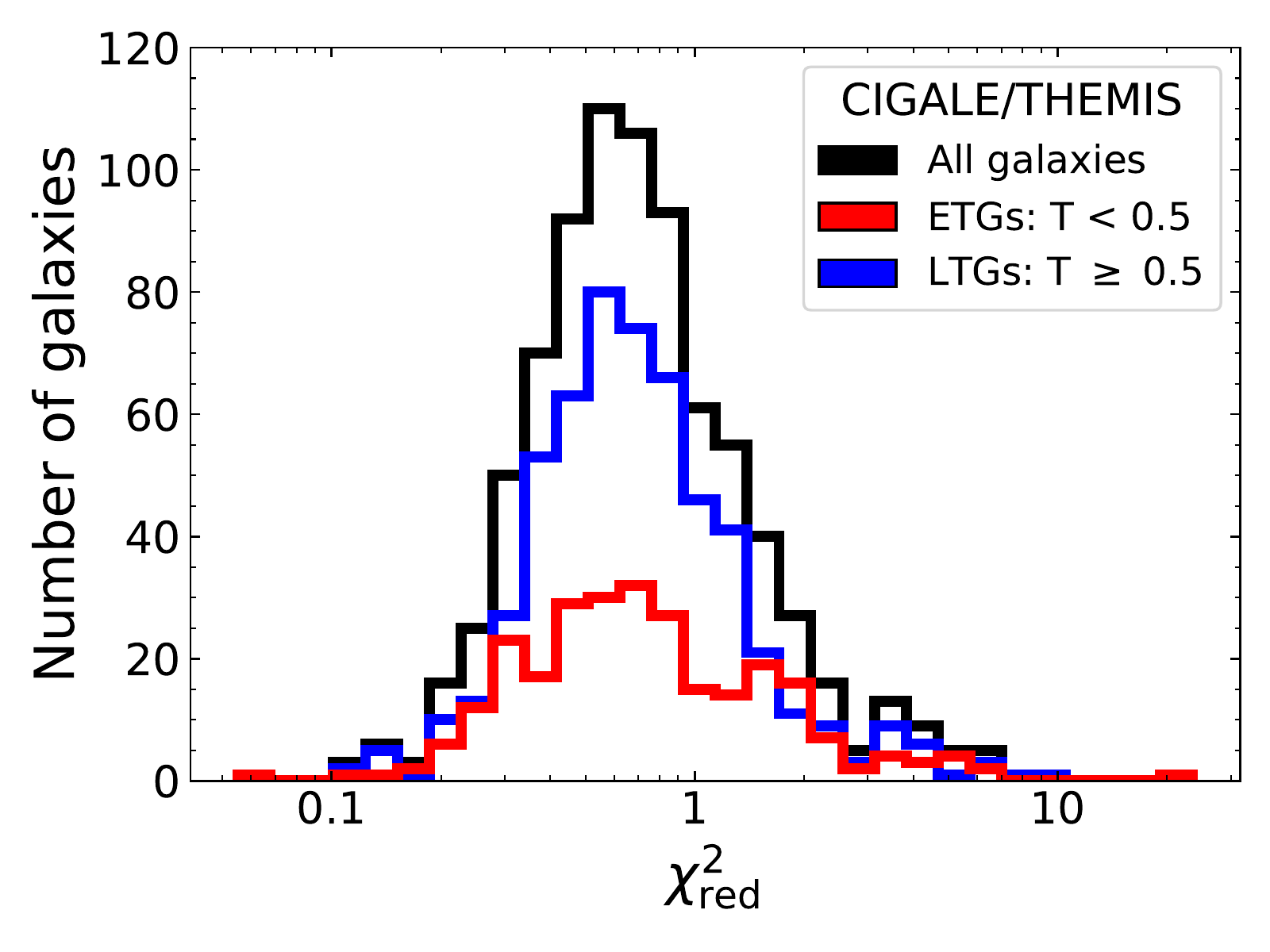}
\caption{Distribution of the reduced $\chi^2$ for the 814 galaxies modelled with \textsc{CIGALE} and with the \textsc{THEMIS} dust model (black line). The distributions for the LTG and ETG subsamples are shown with blue and red lines respectively.}
\label{fig:chisqr}
\end{figure}

\begin{figure*}[t]
\centering
\includegraphics[width=18cm]{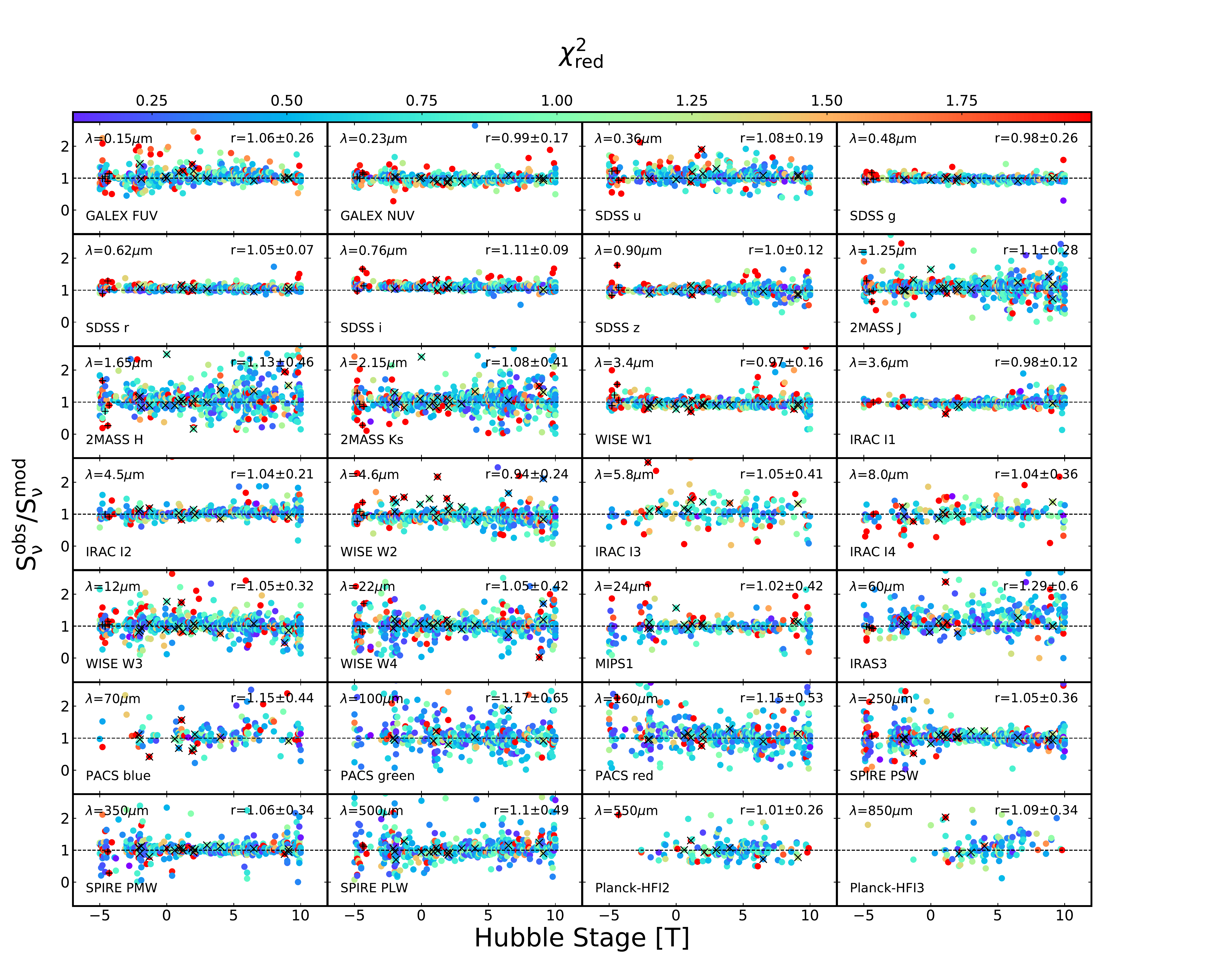}
\caption{Ratios of the observed fluxes to the modelled (\textsc{CIGALE}) fluxes at specific wavelengths, from the FUV ($0.15~\mu$m) to the submm ($850~\mu$m), as a function of Hubble stage ($T$). In the top-left corner of each panel, the wavelength indication (in $\mu$m) is given and in the bottom-left corner the name of the survey/band is provided. The ratios are colour-coded with the $\chi^2_\mathrm{red}$ of the fit as indicated in the colour-bar on the top of the plot. Galaxies hosting an AGN or strong radio-jets are marked with an `X' or a `+' respectively. In the top-right corner of each panel, the mean value of the ratio and the standard-deviation are also provided.}
\label{fig:flux_res}
\end{figure*}

We adopted a similar parameter grid as the one used by \citet{2018arXiv180904088H} to fit the SEDs of 61 galaxies from the KINGFISH sample \citep{2011PASP..123.1347K}. In Table~\ref{tab:param} we give the parameter space used by \textsc{CIGALE} to calculate the SED templates to be fitted to the actual datasets. Two sets of templates were produced, one including the dust parameters given by the \textsc{THEMIS} model and, for comparison, another one including the DL14 dust model characteristics (the latter is briefly discussed in Appendix~\ref{ap:dl14}). In total, 80,041,500 such templates were created with \textsc{CIGALE} running on the high performance cluster of Ghent University. The parameters derived for each galaxy are provided in the DustPedia archive while the mean values, per morphological type, for SFR, $M_\mathrm{star}$, and $M_\mathrm{dust}^\textsc{CIGALE}$ are given in Table~\ref{tab:phys_param}.

\subsubsection{Quality of the fit} \label{subsubsec:mock}

With the current set up of \textsc{CIGALE} we can derive estimates of several parameters such as the SFR, the FUV attenuation ($A_\mathrm{FUV}$), the minimum ISRF intensity ($U_\mathrm{min}$), the stellar mass ($M_\mathrm{star}$), the bolometric luminosity ($L_\mathrm{bolo}$), the dust mass and luminosity ($M_\mathrm{dust}^\textsc{CIGALE}$ and $L_\mathrm{dust}$ respectively), the mass fraction of hydrocarbon solids ($q_\mathrm{hac}$; the PAH abundance $q_\mathrm{PAH}$ in the case of DL14), and the fraction of the dust luminosity coming from PDRs ($\gamma$), to name a few. To examine how well these parameters can be constrained from the multi-wavelength SED fitting that \textsc{CIGALE} performs, as well as the accuracy and precision expected for each parameter, we made use of the \textsc{CIGALE} module that performs a mock analysis. This module creates a mock SED for each galaxy, based on the best fitted parameters, allowing the fluxes to vary within their measured uncertainties. By modelling these mock SEDs with \textsc{CIGALE} we can then retrieve the best set of the mock fitted parameters and compare them with those used as an input. This provides us with a direct measure of how accurately one can retrieve the specific parameters for the specific sample of galaxies. The results of the mock analysis can be found in Appendix~\ref{ap:themis}.

To explore the overall quality of the fits to the observations we examine the distribution of the reduced $\chi^2$ values ($\chi^2_\mathrm{red}$; the $\chi^2$ values divided by the number of observations minus the number of free parameters). The $\chi^2_\mathrm{red}$ distribution is shown in Fig.~\ref{fig:chisqr} (black line). The $\chi^2_\mathrm{red}$ distributions for the two main morphological classes of galaxies in the DustPedia sample [ETGs ($T<0.5$) and LTGs ($T\ge0.5$)], are also shown (red and blue lines respectively). For the full sample modelled with \textsc{CIGALE} (814 galaxies) we find that the median value of the histogram is 0.66, while it gets to 0.67 when only considering the 546 LTGs and drops down again to 0.66 for the 268 ETGs. Out of the 814 modelled galaxies, there are 60 ($\sim7\%$) with $\chi^2_\mathrm{red}>2$ and only 22 ($\sim3\%$) with $\chi^2_\mathrm{red}>4$. Similar distributions for the $\chi^2_\mathrm{red}$ are obtained when using the DL14 dust model (see Fig.~\ref{fig:chisqr_dl14} in Appendix~\ref{ap:dl14}), although the values are slightly lower compared to \textsc{THEMIS}. A possible explanation is that \textsc{THEMIS} has on average a flatter FIR-submm slope ($\beta = 1.79$) than the DL14 model ($\beta = 2$). With a distribution of ISRF such as the one we assume, it is always possible to fit an observed slope flatter than the slope of the intrinsic grain properties, by just adding colder temperatures. This means that the model with the highest $\beta$ (DL14) has the highest fitting flexibility, and will thus result in lower $\chi^2_\mathrm{red}$ values.

\begin{figure}[t]
\centering
\includegraphics[width=9cm]{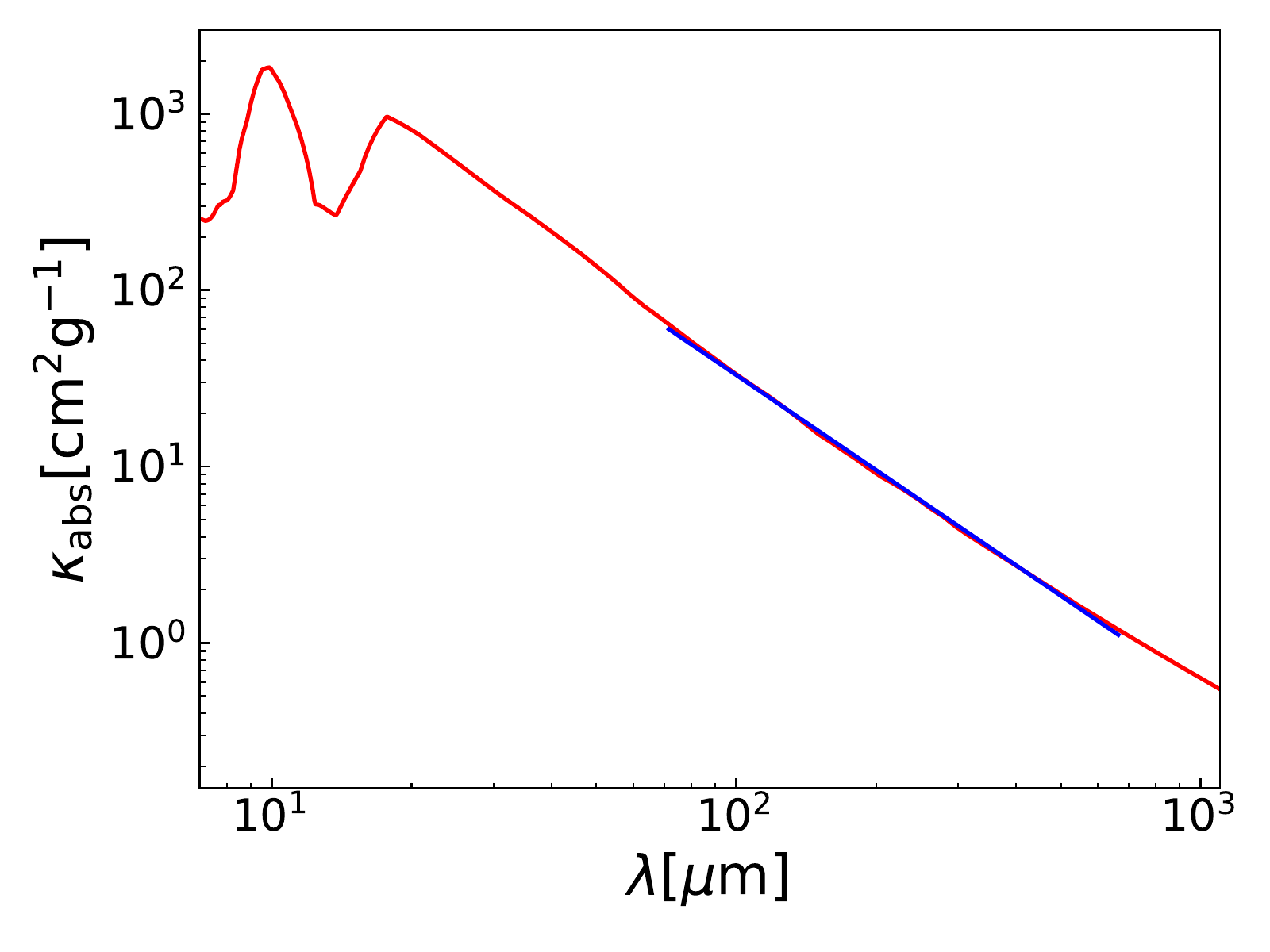}
\caption{The average absorption coefficient calculated within the \textsc{THEMIS} dust model (red line). For the wavelength range $70 \le \lambda/\mu \text{m} \le 700$ we have approximated the extinction law with a power-law (blue line) so that it can be used in the MBB calculations (see text for more details).}
\label{fig:kabs_j17}
\end{figure}

As a further check of the goodness of the fit to the observations we examine how the ratios of the observed-to-modelled flux densities compare for galaxies of different Hubble stages. The model flux densities in each waveband were calculated by \textsc{CIGALE}. A systematic trend in these ratios in a given band could help reveal potential weaknesses of our modelling. This is shown in Fig.~\ref{fig:flux_res} with the ratios at wavelengths ranging from the FUV ($0.15~\mu$m; top-left panel) to the submm ($850~\mu$m; bottom-right panel) with the wavelength indicated in the top-left corner in each panel. Each point represents a galaxy colour-coded with its $\chi^2_\mathrm{red}$. Overall we see that, despite the large scatter in some cases, the ratio of the observed-to-modelled flux densities stays around unity (horizontal dashed line) indicating that \textsc{CIGALE} is able to adequately fit the SED of the galaxy in the full wavelength range considered. The $\chi^2_\mathrm{red}$ values (as indicated with the different colours) show a general picture where many ETGs and irregular galaxies (the two extremes in the \textit{x}-axis) are either on the higher end (red colour) or the lower end (blue colour) on the $\chi^2_\mathrm{red}$ scale while the galaxies with intermediate Hubble stages get $\chi^2_\mathrm{red}$ values closer to one (cyan and green colours). 

For each waveband we have calculated (and presented in the top-right corner in each panel) the mean value of the ratio as well as the standard-deviation. We see that there are 14 wavebands which show deviations to the mean of less than or equal to $5\%$ (0.23, 0.48, 0.62, 0.90, 3.4, 3.6, 4.5, 5.8, 8.0, 12, 22, 24, 250, and 550~$\mu$m), 8 with deviations larger than $5\%$ and less than or equal to $10\%$ (0.15, 0.36, 1.25, 2.15, 4.6, 350, 500 and 850~$\mu$m) and the remaining 6 with deviations larger than $10\%$ (0.76, 1.65, 60, 70, 100, and 160~$\mu$m). In all six cases with the largest deviations from unity ($>10\%$) the model under-predicts the observed fluxes. With the exception of the SDSS \textit{i}-band (0.76~$\mu$m) with a scatter of 0.09, the rest of the bands show large scatter ($>0.4$) and especially for Hubble stages $T<0$ and $T>5$ which also drives the underestimation of the modelled fluxes. At submm wavelengths and especially at 500 and 850~$\mu$m there seems to be a mild trend of increasing of the observed-to-modelled flux ratios with increasing Hubble stage. Such a trend has already been reported in other studies \citep{2014A&A...565A.128C} with a possible interpretation being the submm excess observed in low-metallicity systems \citep{2009A&A...508..645G, 2010A&A...518L..55G, 2014A&A...565A.128C}. 

\begin{figure}[t]
\centering
\includegraphics[width=9cm]{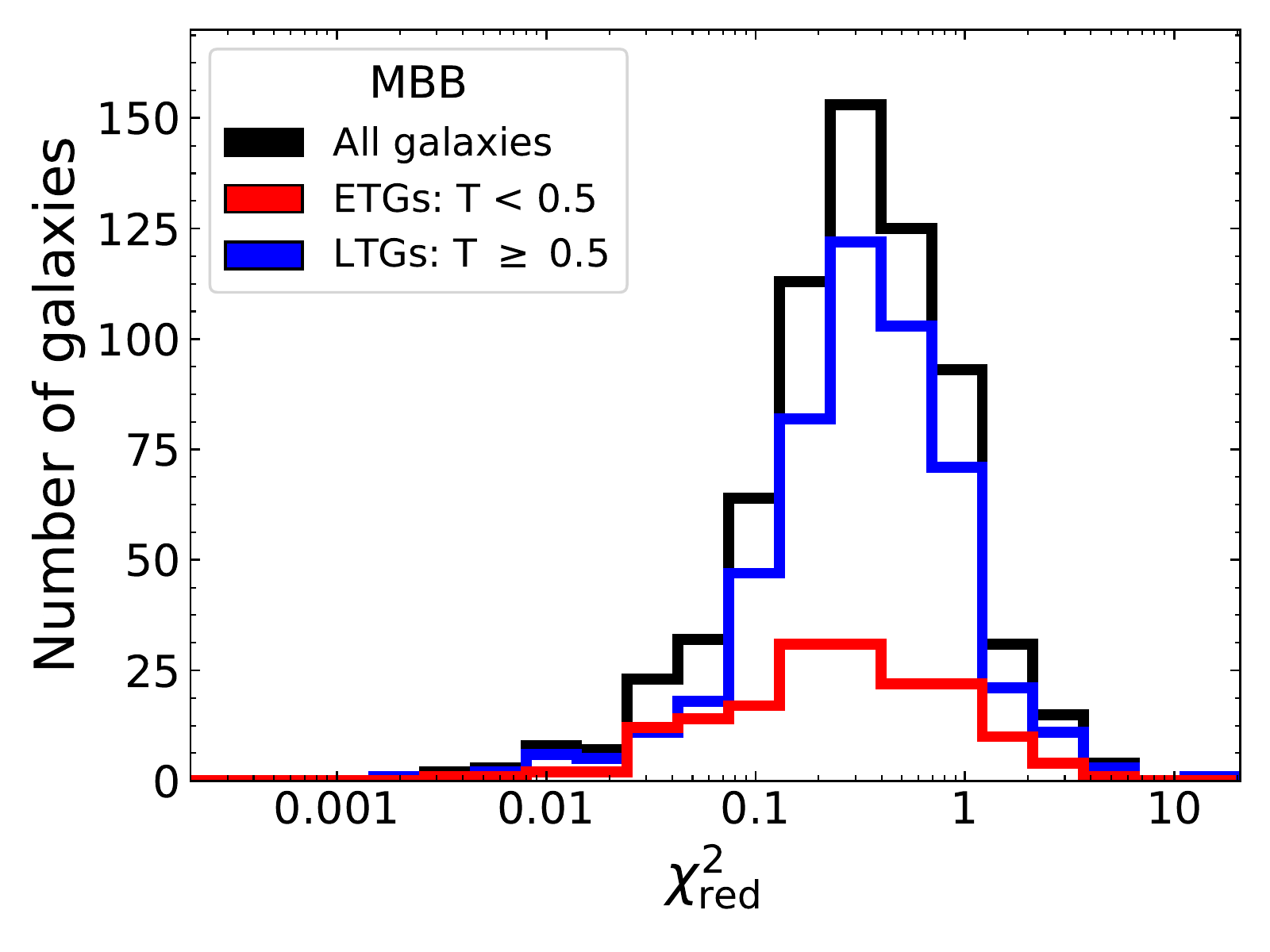}
\caption{Distribution of the reduced $\chi^2$ of the 678 galaxies modelled with a MBB scaled to the \textsc{THEMIS} dust mosel (black line). The distributions for the LTG and the ETG subsamples are also shown (blue and red lines respectively).}
\label{fig:chisqrmbb}
\end{figure}

\begin{figure*}[t]
\centering
\includegraphics[width=18cm]{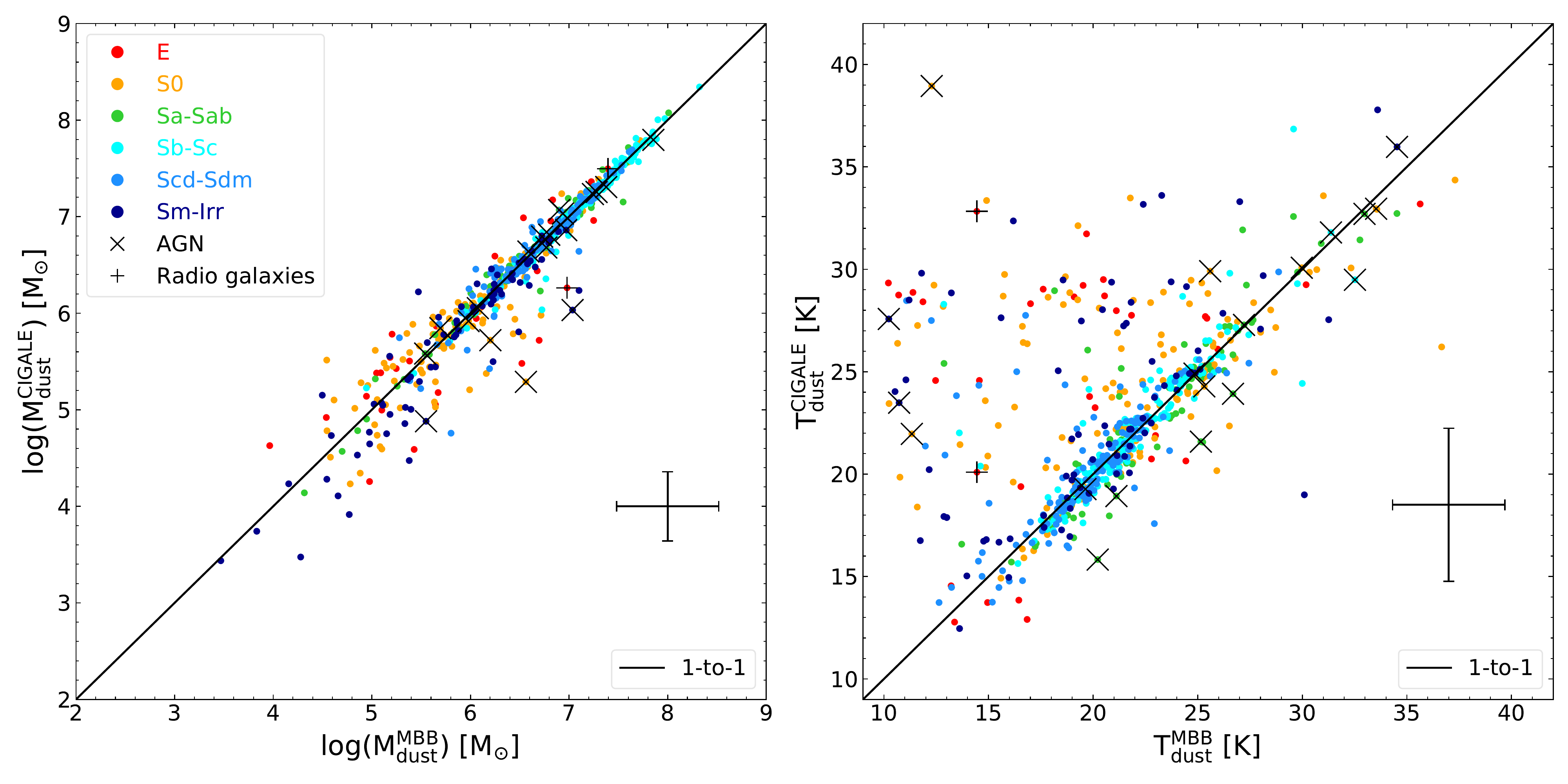}
\caption{Comparison of $M_\mathrm{dust}$ (left panel) and $T_\mathrm{dust}$ (right panel), as derived from \textsc{CIGALE} (\textit{y}-axis) and MBB (\textit{x}-axis) modelling. The points are colour-coded according to the morphological type, while `+' and `X' symbols indicate strong radio jet galaxies and AGNs respectively (see the inset in the left panel for the explanation of the different colours and symbols). The solid black line indicates the one-to-one relation. The average uncertainty is indicated at the lower right corner in each plot.}
\label{fig:comp_mbb}
\end{figure*}

\subsection{Fitting the FIR with Modified Black-Bodies}\label{MBB}

The emission from dust in thermal equilibrium with the radiation field can be approximated using modified black-bodies with flux densities given by:

\begin{equation} \label{eq:mbb}
S(\lambda, T_\mathrm{dust}^\mathrm{MBB}) \ \propto \ \lambda^{-\beta}B(\lambda, T_\mathrm{dust}^\mathrm{MBB}) \, ,
\end{equation}

\noindent where $\beta$ is the grain emissivity index, usually taking values between 1 and 2 \citep{1983QJRAS..24..267H, 1984ApJ...285...89D, 2012A&A...540A..54B, 2013MNRAS.428.1880A}, and $B(\lambda, T_\mathrm{dust}^\mathrm{MBB})$ the Planck function at a given temperature $T_\mathrm{dust}^\mathrm{MBB}$.

Integrating the flux density over a certain wavelength range provides the luminosity emitted by the dust at those wavelengths. For consistency we calculate the dust luminosity in the wavelength range from 8 to 1000~$\mu$m as usually used \citep{1998ARA&A..36..189K}. Assuming an opacity $\kappa\left( \lambda \right)$ for the average dust grain mix, the dust mass can then be derived by: 

\begin{equation} \label{eq:mdust}
M_\mathrm{dust}^\mathrm{MBB} \ = \ \frac{D^2}{\kappa\left( \lambda \right)} \ \frac{S(\lambda, T_\mathrm{dust}^\mathrm{MBB})}{B(\lambda, T_\mathrm{dust}^\mathrm{MBB})} \, ,
\end{equation}

\noindent \citep{1983QJRAS..24..267H} with $D$ being the distance to the galaxy (in Mpc). 

The opacity is usually approximated by a power-law over the infrared wavelengths \citep[see, e.g.,][]{2004A&A...425..109A, 2018ARA&A..56..673G}. In our case, in order to be consistent with the grain physics of the \textsc{THEMIS} model, we have fitted a power-law to the average opacity inferred by \textsc{THEMIS} (see Fig.~\ref{fig:kabs_j17}), in the wavelength range $70 \le \lambda/\mu\mathrm{m} \le 700$. We find that the opacity scales with wavelength as:

\begin{equation} \label{eq:kabs}
\kappa\left( \lambda \right) \ = \ \kappa_{250} \ \left(250/\lambda\right)^{1.790} \, ,
\end{equation}

\noindent with $\lambda$ given in $\mu$m and $\kappa_{250} = 6.40~\mathrm{cm}^2~\mathrm{g}^{-1}$. 

We modelled the DustPedia galaxies with a single MBB using data at wavelengths $\lambda \ge 100~\mu$m, i.e., every available observation among the IRAS ($100~\mu$m), PACS (100, 160~$\mu$m), MIPS ($160~\mu$m), SPIRE (250, 350, 500~$\mu$m) and \textit{Planck} (350, 550~$\mu$m) wavebands. We avoided using fluxes below $100~\mu$m so that the fitted SED is not polluted by emission from dust grains in non-thermal equilibrium as well as fluxes above $550~\mu$m so that possible contamination from synchrotron and free-free emission from low-luminosity radio galaxies, is not included. As in the case of \textsc{CIGALE}, a $10\%$ uncertainty was added quadratically to the measured flux uncertainties (see Sect.~\ref{subsubsec:par_space}). The fit was made using standard $\chi^2$ minimization techniques (Levenberg-Marquardt) allowing the dust temperature ($T_\mathrm{dust}^\mathrm{MBB}$) to range between 10 and 40~K. The SED of the MBB was convolved to each filter's RSRF. In the case of the SPIRE bands the RSRF for extended emission was used. An estimate of the uncertainty on the derived parameters is provided by performing a bootstrap analysis to our datasets by fitting 1000 SEDs for each galaxy. The mock fluxes are randomly drawn from a Gaussian distribution centered on the observed flux and with a standard-deviation identical to those observed. The uncertainty assigned to each of the two parameters ($T_\mathrm{dust}^\mathrm{MBB}$ and $M_\mathrm{dust}^\mathrm{MBB}$) is then defined as the standard-deviation of the 1000 values derived from this procedure.

Out of the 875 DustPedia galaxies, 802 have at least three reliable measurements (with no major flags associated) in the wavelength range under consideration (100-600~$\mu$m) and could be fitted with a MBB. Out of these 802 galaxies, only 678 galaxies fulfilled our temperature boundary conditions (10-40~K) and gave a reasonable fit. The parameters derived for each galaxy are provided in the DustPedia archive while the mean values, per morphological type, for $M_\mathrm{dust}^\mathrm{MBB}$, and $T_\mathrm{dust}^\mathrm{MBB}$ are given in Table~\ref{tab:phys_param}.

As in \textsc{CIGALE}, we explore the overall quality of the fits performed to the observations by examining the distribution of the $\chi^2_\mathrm{red}$ values. The $\chi^2_\mathrm{red}$ distribution is shown in Fig.~\ref{fig:chisqrmbb}. We find that the median value of the distribution (for the 678 modelled galaxies) is at 0.31 while it gets to 0.32 when only considering the 506 LTGs and drops down to 0.25 for the 172 ETGs. Out of the 678 modelled galaxies there are 22 ($\sim3\%$) with $\chi^2_\mathrm{red}>2$ and only four galaxies with $\chi^2_\mathrm{red}>4$. Here, the small values of $\chi^2_\mathrm{red}$ (much smaller than unity) indicate that the model is `over-fitting' the data. This is mainly due to the fact that the number of available observations is, in many cases, small, but also that the noise assigned to the fluxes is sufficiently large to allow for a poorly constrained model. 

\subsection{Comparison between \textsc{CIGALE} and MBB}

While \textsc{CIGALE} does not provide a direct estimate for $T_\mathrm{dust}$, it allows us to approximate it using the strength of the ISRF parametrized by $U_\mathrm{min}$. Assuming that dust is heated by an ISRF with a Milky-Way like spectrum \citep{1983A&A...128..212M}, we can approximate $T_\mathrm{dust}$ by:

\begin{equation} \label{eq:Tdust}
T_\mathrm{dust}^\textsc{CIGALE} \ = T_\mathrm{o} \ U_\mathrm{min}^{(1/(4+\beta))} \, ,
\end{equation}   

\noindent \citep{2012ApJ...756..138A}. Here, $U_\mathrm{min}$ is the minimum ISRF level heating the diffuse dust, $T_\mathrm{o} = 18.3$~K is the dust temperature measured in the solar neighbourhood, and $\beta$ is the dust emissivity index, which, for the \textsc{THEMIS} dust model, gets the value of 1.79 (see Sect.~\ref{MBB}). The values of $T_\mathrm{dust}^\textsc{CIGALE}$ derived for each galaxy are provided in the DustPedia archive while the mean values, per morphological type, are given in Table~\ref{tab:phys_param}.

Having derived the dust masses and temperatures using the two methods described above (\textsc{CIGALE} and MBB) we can directly compare them for the 678 galaxies in common in the two sub-samples. We do so in Fig.~\ref{fig:comp_mbb}, with the comparison of $M_\mathrm{dust}$ in the left panel and that of $T_\mathrm{dust}$ in the right panel. In each panel the symbols are colour-coded according to the morphological class.

The two different methods of estimating the dust properties compare fairly well, with the scatter, generally, increasing for less dusty objects (ETGs). This is evident in the dust masses (left panel of Fig.~\ref{fig:comp_mbb}), with most of the deviant cases showing MBB dust masses higher than those derived by \textsc{CIGALE}, but it becomes more prominent in the dust temperatures (right panel of Fig.~\ref{fig:comp_mbb}) with \textsc{CIGALE} systematically estimating higher values compared to MBB for these galaxies. Some of the galaxies identified as AGNs and strong radio-sources are amongst the outliers indicating that the models could not adequately fit the observations, though these are only a few cases. For the majority of the most deviant cases, a visual inspection of the SEDs reveals that this is, mainly, due to a combination of two effects. First, the FIR-submm measurements, especially for the ETGs, come with large uncertainties due to the low level of emission. This allows \textsc{CIGALE}, which is constrained from a large multi-wavelength dataset, for a larger `flexibility' in fitting this part of the SED, in most cases `over-weighting' MIR and FIR against submm measurements. This results in a bias favoring high temperatures for \textsc{CIGALE} compared to MBB modelling where only the few FIR-submm data are fitted. Furthermore, a single MBB component fitted to the FIR-submm fluxes is only sensitive to the colder dust, missing a large fraction of dust heated to warmer temperatures. These are the effects that we see in Fig.~\ref{fig:comp_mbb} but also looking at the mean values of the dust temperature in Table~\ref{tab:phys_param} where we see that for galaxies with $0<T<8$ there is a very good agreement between the two methods, well within the uncertainties, while the differences progressively become larger for earlier- and later-type galaxies. Overall, we conclude that the dust temperatures derived from \textsc{CIGALE} are more accurate than those from the MBB fitting. In spite of that, we plan to further investigate the discrepancy between the two methods in future papers. The dust temperatures obtained by \textsc{CIGALE} ($T_\mathrm{dust}^\textsc{CIGALE}$) as a function of morphological type are discussed in more detail in Sect.~\ref{sec:dsr}. 

\section{Physical parameters as a function of the Hubble stage} \label{sec:dsr}

The abundance of dust and gas in galaxies can reveal valuable information about their star-formation cycle and chemical evolution. The local environment is expected to have a significant role in that regard, affecting the chemical evolution and the ISM content in individual galaxies \citep[e.g.,][]{2004A&A...422..941C, 2006MNRAS.373..469B, 2009ApJ...697.1811F, 2010ApJ...721..193P, 2012MNRAS.423.1277D}. Davies et al. (2019, to be submitted), investigated the role of the environment for a subset of the DustPedia galaxies (grouped into field and cluster members) by examining their stellar, dust and gas content. They found that the physical properties of galaxies of the same morphological type do not vary significantly with environment suggesting that the intrinsic properties of the galaxies are determined, mainly, by their internal physical processes.

\begin{figure}[t]
\centering
\includegraphics[width=9cm]{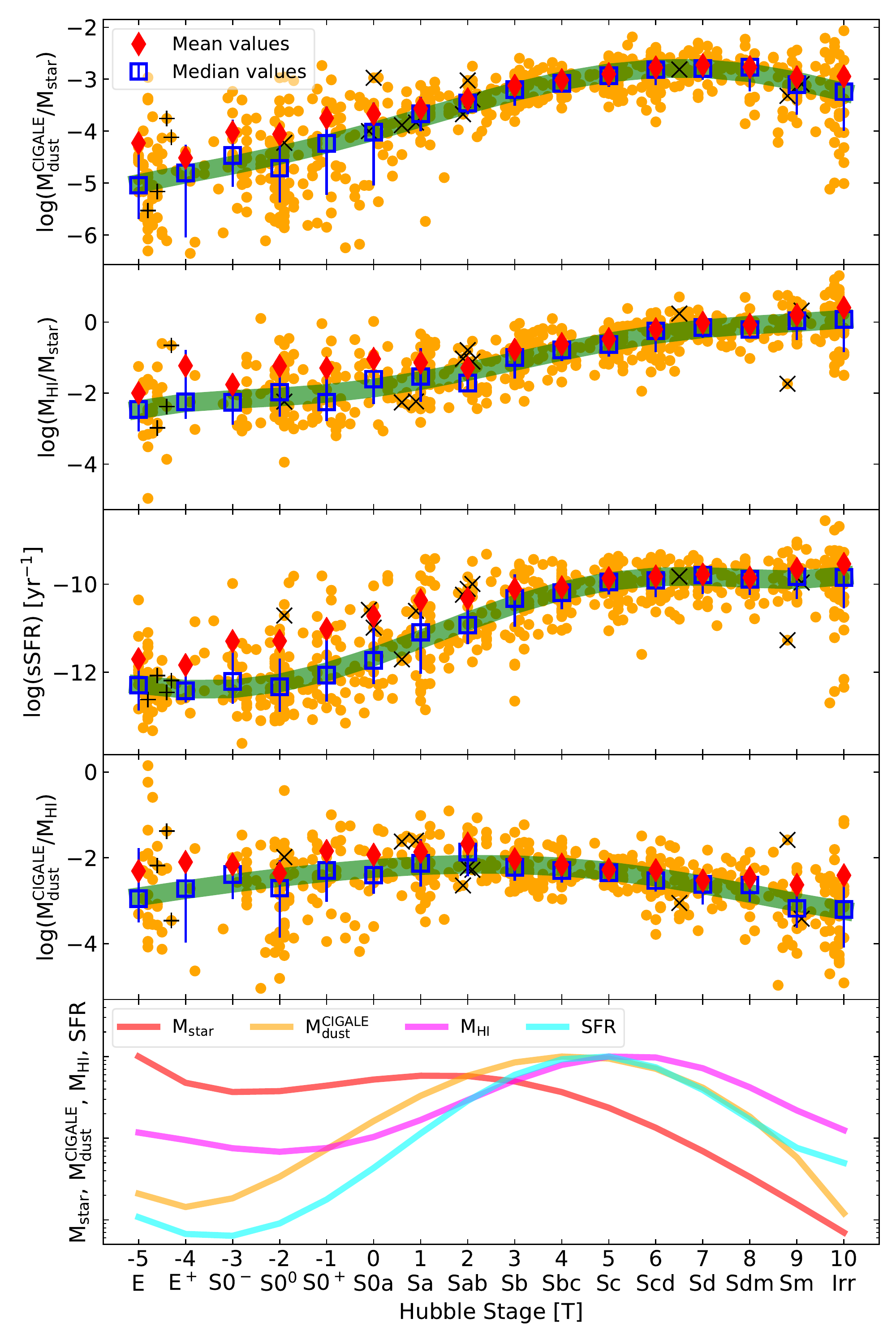}
\caption{From top to bottom (first four panels): $M_\mathrm{dust}^\textsc{CIGALE}$/$M_\mathrm{star}$, $M_\mathrm{HI}$/$M_\mathrm{star}$, sSFR and  $M_\mathrm{dust}^\textsc{CIGALE}$/$M_\mathrm{HI}$ as a function of Hubble stages ($T$). In each panel orange circles are individual galaxies, red diamonds are the mean values for each morphological bin, while blue squares are the median values. Error bars bracket the range between the 16th and 84th percentiles from the median. The thick green curves are 5th order polynomial regressions to the median values (see Table~\ref{tab:spline} for the polynomial regression parameters). In each of these panels `+' and `X' symbols indicate strong radio jet galaxies and AGNs respectively. The last panel shows the variation in $M_\mathrm{star}$, $M_\mathrm{dust}^\textsc{CIGALE}$, $M_\mathrm{HI}$, and SFR (red, orange, magenta, and cyan lines respectively) with morphology as fitted with a 5th order polynomial regression through the median values (see Table~\ref{tab:spline}). $M_\mathrm{star}$, $M_\mathrm{dust}^\textsc{CIGALE}$, $M_\mathrm{HI}$, and SFR have been normalized to the maximum value obtained from each polynomial regression by: $3.9\times10^{10}$~M$_{\odot}$, $1.2\times10^7$~M$_{\odot}$, $2.6\times10^9$~M$_{\odot}$ and 1.08~M$_{\odot}$/yr respectively.}
\label{fig:dsr}
\end{figure}

Although the general picture that local ETGs are poor in ISM (dust and gas) compared to less evolved galaxies is well accepted \citep[e.g.,][]{2003A&A...405....5B, 2004A&A...422..941C, 2014A&A...564A..66B}, a more detailed analysis is needed on how the ISM content varies among galaxies of different morphological stages. The DustPedia sample is ideal to carry out this analysis since it contains the most complete, to date, collection of local galaxies spanning the full range of morphologies with sufficient multi-wavelength coverage so that SED fitting analysis can be applied in a uniform way. Furthermore, measurements of the atomic hydrogen mass ($M_\mathrm{HI}$) for a significant fraction of the DustPedia galaxies (711 out of 814 galaxies; see Table~\ref{tab:phys_param}) allow us to investigate how the gas content in galaxies varies with morphology. 

\begin{table*}[t]
\caption{Mean values of $M_\mathrm{dust}^\textsc{CIGALE}$, $M_\mathrm{HI}$, and SFR normalized to the $M_\mathrm{star}$, as well as the $M_\mathrm{dust}^\textsc{CIGALE}$/$M_\mathrm{HI}$ ratio together with their associated standard deviation for different morphological bins.}
\begin{center}
\scalebox{1.0}{
\begin{tabular}{c|ccc||c}
\hline 
\hline 
$T$ & 
$\log\left(\left\langle M_\mathrm{dust}^\textsc{CIGALE}/M_\mathrm{star}\right\rangle\right)$ & 
$\log\left(\left\langle M_\mathrm{HI}/M_\mathrm{star}\right\rangle\right)$ &
$\log\left(\left\langle SFR/M_\mathrm{star}\right\rangle\right)$ &
$\log\left(\left\langle M_\mathrm{dust}^\textsc{CIGALE}/M_\mathrm{HI}\right\rangle\right)$\\
\hline
    -5 & -4.23 $\pm$ 0.33 & -2.00 $\pm$ 0.08 & -11.70 $\pm$ 0.42 & -2.31 $\pm$ 0.45 \\
    -4 & -4.51 $\pm$ 0.52 & -1.23 $\pm$ 0.05 & -11.84 $\pm$ 0.48 & -2.10 $\pm$ 0.41 \\
    -3 & -4.02 $\pm$ 0.42 & -1.76 $\pm$ 0.05 & -11.29 $\pm$ 0.57 & -2.15 $\pm$ 0.34 \\
    -2 & -4.06 $\pm$ 0.35 & -1.24 $\pm$ 0.09 & -11.29 $\pm$ 0.17 & -2.36 $\pm$ 0.20 \\
    -1 & -3.75 $\pm$ 0.19 & -1.29 $\pm$ 0.08 & -11.02 $\pm$ 0.18 & -1.84 $\pm$ 0.15 \\
    0  & -3.67 $\pm$ 0.30 & -1.04 $\pm$ 0.16 & -10.72 $\pm$ 0.18 & -1.93 $\pm$ 0.25 \\
    1  & -3.57 $\pm$ 0.15 & -1.14 $\pm$ 0.11 & -10.37 $\pm$ 0.14 & -1.86 $\pm$ 0.11 \\
    2  & -3.39 $\pm$ 0.15 & -1.29 $\pm$ 0.09 & -10.32 $\pm$ 0.16 & -1.67 $\pm$ 0.07 \\
    3  & -3.13 $\pm$ 0.13 & -0.78 $\pm$ 0.11 & -10.10 $\pm$ 0.13 & -2.04 $\pm$ 0.06 \\
    4  & -3.02 $\pm$ 0.14 & -0.62 $\pm$ 0.10 & -10.07 $\pm$ 0.14 & -2.16 $\pm$ 0.07 \\
    5  & -2.90 $\pm$ 0.15 & -0.48 $\pm$ 0.12 & -9.86 $\pm$ 0.13  & -2.28 $\pm$ 0.08 \\
    6  & -2.77 $\pm$ 0.19 & -0.20 $\pm$ 0.11 & -9.83 $\pm$ 0.14  & -2.29 $\pm$ 0.12 \\
    7  & -2.73 $\pm$ 0.21 & -0.02 $\pm$ 0.13 & -9.78 $\pm$ 0.13  & -2.53 $\pm$ 0.11 \\
    8  & -2.77 $\pm$ 0.19 & -0.07 $\pm$ 0.10 & -9.85 $\pm$ 0.12  & -2.46 $\pm$ 0.13 \\
    9  & -2.96 $\pm$ 0.30 &  0.20 $\pm$ 0.15 & -9.65 $\pm$ 0.14  & -2.63 $\pm$ 0.36 \\
    10 & -2.95 $\pm$ 0.39 &  0.41 $\pm$ 0.17 & -9.54 $\pm$ 0.19  & -2.41 $\pm$ 0.39 \\
\hline
    $\left[-5.0,-3.5\right)$ & -4.29 $\pm$ 0.36 & -1.66 $\pm$ 0.06 & -11.73 $\pm$ 0.43 & -2.24 $\pm$ 0.44 \\
    $\left[-3.5,0.5\right)$  & -3.88 $\pm$ 0.30 & -1.23 $\pm$ 0.11 & -11.06 $\pm$ 0.21 & -2.05 $\pm$ 0.20 \\
    $\left[0.5,2.5\right)$   & -3.48 $\pm$ 0.15 & -1.20 $\pm$ 0.10 & -10.35 $\pm$ 0.15 & -1.77 $\pm$ 0.09 \\
    $\left[2.5,5.5\right)$   & -3.00 $\pm$ 0.14 & -0.60 $\pm$ 0.11 & -9.99  $\pm$ 0.13 & -2.15 $\pm$ 0.07 \\
    $\left[5.5,8.5\right)$   & -2.76 $\pm$ 0.20 & -0.11 $\pm$ 0.11 & -9.82  $\pm$ 0.13 & -2.38 $\pm$ 0.12 \\
    $\left[8.5,10.0\right]$  & -2.95 $\pm$ 0.36 &  0.34 $\pm$ 0.17 & -9.58  $\pm$ 0.18 & -2.48 $\pm$ 0.39 \\
\hline \hline
\end{tabular}}
\label{tab:dsr}
\end{center}
\end{table*}

In the first three panels of Fig.~\ref{fig:dsr} we investigate how the specific mass of dust ($M_\mathrm{dust}^\textsc{CIGALE}$/$M_\mathrm{star}$) and atomic gas ($M_\mathrm{HI}$/$M_\mathrm{star}$), as well as the sSFR (SFR/$M_\mathrm{star}$), vary with the morphology of the galaxy. In each panel the orange circles show the values of individual galaxies while the red diamonds and blue squares show the mean (also given in Table~\ref{tab:dsr}) and the median values for each morphological bin respectively. The fourth panel from top shows the variation of the dust-to-atomic-gas mass ratio as a function of morphological type. We see that $M_\mathrm{dust}^\textsc{CIGALE}$/$M_\mathrm{star}$, $M_\mathrm{HI}$/$M_\mathrm{star}$ and SFR/$M_\mathrm{star}$ vary by about two orders of magnitude whereas $M_\mathrm{dust}^\textsc{CIGALE}$/$M_\mathrm{HI}$ varies by one order of magnitude.
Finally, the bottom panel shows the change in $M_\mathrm{star}$, $M_\mathrm{dust}^\textsc{CIGALE}$, $M_\mathrm{HI}$, and SFR with Hubble stage as a 5th order polynomial regression through the median values per Hubble stage bin.

From the bottom panel of Fig.~\ref{fig:dsr}, we see that the stellar mass (red line) takes its maximum value for E type galaxies and varies slightly for galaxies with $T<2$. For galaxies with $T>2$, a sharp drop in stellar mass (of about two orders of magnitude) is observed. The HI mass (magenta line) varies slightly for galaxies with $\mathrm{T}<2$ (in a similar way as stellar mass), followed by a rise (of about one order of magnitude) and then a drop for galaxies with $\mathrm{T}>5$. The dust mass (orange line) and the SFR (cyan line) vary in a similar way with a continuous increase for ETGs and a decrease for later-type galaxies with a peak value for Sc type galaxies. However, the real variation of the ISM content between different galaxies can only be appreciated if we consider galaxies of the same stellar mass. This is what we present in the first three panels by normalizing $M_\mathrm{dust}^\textsc{CIGALE}$, $M_\mathrm{HI}$, and SFR, to the stellar mass of each galaxy.

A continuous increase of the dust mass (about two orders of magnitude, on average) is observed for galaxies with Hubble stages from -5 up to around 7, where $M_\mathrm{dust}^\textsc{CIGALE}$/$M_\mathrm{star}$ peaks, and then drops at larger Hubble stages (top panel). The continuous increase of the ratio is mainly due to the sharp increase in dust mass from $T=-5$ to $T=5$ while from $T=5-7$ it is mainly the drop in the stellar mass that keeps the ratio increasing. Then, from $T=7$ and beyond, it is mainly the dust mass that takes over again with a sharp drop (see the orange and red lines in the bottom panel for the dust and the stellar mass changes). 

The specific mass of the atomic gas content, on the other hand, shows a relatively flat behaviour for ETGs (being roughly constant and around 0.01 for galaxies with $T\sim2$). This is mainly because both the stellar mass and the HI mass show similar trends for these types of galaxies. Meanwhile, the decrease in stellar mass and a mild increase of the atomic gas mass for galaxies with $2<T<7$ is the main driver of the increase of the $M_\mathrm{HI}$/$M_\mathrm{star}$ ratio of about one order of magnitude. The sharp decrease in the stellar mass for galaxies beyond $T>7$ then compensates the drop in the gas mass keeping the $M_\mathrm{HI}$/$M_\mathrm{star}$ ratio increasing (to about unity) but with a slower rate. 

The sSFR stays roughly constant for earlier-type galaxies ($T<-2$). This is because, as we see from the bottom panel, $M_\mathrm{star}$ and SFR roughly follow the same trend. For galaxies up to $T=5$ though, a sharp increase in SFR and a mild decrease in $M_\mathrm{star}$ (cyan and red lines respectively in the bottom panel) results in a sharp increase in sSFR. For later morphological types, $M_\mathrm{star}$ and SFR follow similar trends resulting in a roughly constant sSFR. The dust-to-gas mass ratio obtains its maximum value around $T=2$. The rise in the earlier types of galaxies comes from the sharp increase in the dust mass, compared to the gas mass which roughly stays constant. Beyond $T=5$, both the dust and the gas mass drop but it is the dust mass that drops faster which drives the slow drop in the $M_\mathrm{dust}^\textsc{CIGALE}$/$M_\mathrm{HI}$ ratio. This behaviour is similar to the results previously presented for the atomic gas mass as a function of morphology in \citet{2007ApJ...663..866D} and \citet{2012A&A...540A..52C} for SINGS \citep{2003PASP..115..928K} and Herschel Reference Survey \citep[HRS,][]{2010PASP..122..261B} galaxies respectively. In these studies, the ratio $M_\mathrm{dust}$/$M_\mathrm{HI}$ peaks for Sab galaxies (around $T=2$) and decreases when going either to ETGs or irregular galaxies.

\begin{figure}[t]
\centering
\includegraphics[width=9cm]{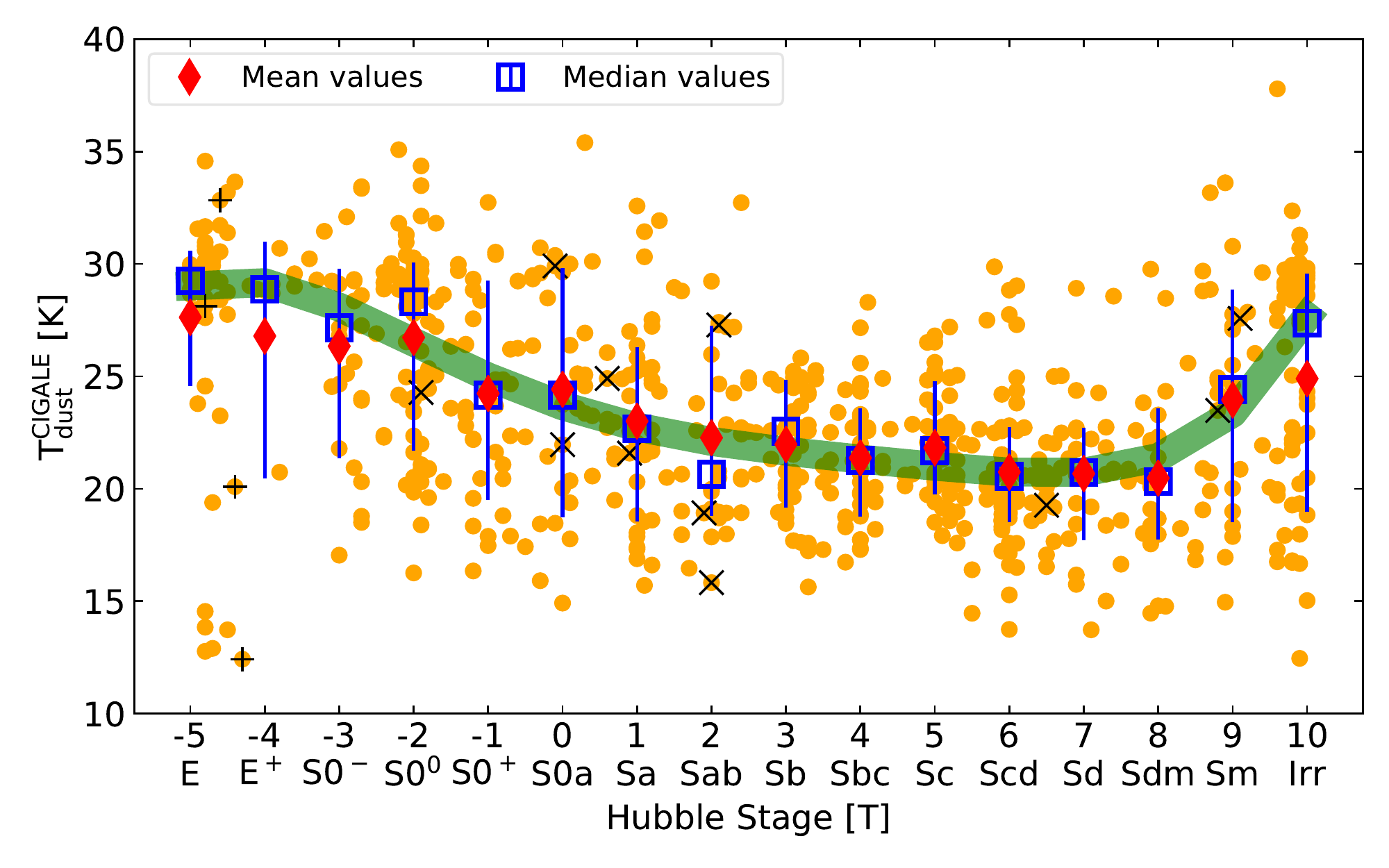}
\caption{Dust temperature, for galaxies of different Hubble stages, as derived from \textsc{CIGALE}. Orange circles are individual galaxies, red diamonds are the mean values for each morphological bin, while blue open squares are the median values. Error bars bracket the range between the 16th and 84th percentiles from the median. The thick green curve is a 5th order polynomial regression to the median values (see Table~\ref{tab:spline}). The `+' and `X' symbols indicate strong radio jet galaxies and AGNs respectively.}
\label{fig:tdust}
\end{figure}

In Fig.\ref{fig:tdust} we show how dust temperature, as obtained by \textsc{CIGALE}, varies with morphology. Although a large scatter is present in each morphological bin, a clear trend is evident with ETGs heating the dust up to higher temperatures ($\sim30$~K) compared to LTGs where a drop in dust temperature (by $\sim10$~K) is observed. These results compare fairly well with the findings of \citet{2011ApJ...738...89S} for 10 ETGs from KINGFISH, which found an average dust temperature of 30~K. A sharp rise in temperature (back to $\sim30$~K) is then seen for Sm-Irr galaxies. Here, we basically see the effect of the intense ISRF seen in ETGs being very efficient in heating the dust up to high temperatures. This is easy to achieve in ETGs since dust, in most cases, is found in the very center of the galaxies where the ISRF intensity is very strong. In LTGs with $0<T<8$, on the other hand, dust found in the disk of the galaxies is distributed in a more diffuse way, away from the heating sources, keeping the dust at low temperatures \citep[see, e.g.,][]{2012A&A...543A..74X}. For Sm-Irr galaxies, where merging processes shape the morphology and triggers star-formation activity, dust is found in the vicinity of star-forming sites giving rise to the higher dust temperatures \citep{2010A&A...518L..61B, 2015A&A...582A.121R}. For these galaxies a potential submm excess is detected (500 and 850$~\mu$m residuals in Fig.\ref{fig:flux_res}). This effect is expected to account for some extra amount of cold dust \citep{2009A&A...508..645G, 2010A&A...518L..55G, 2014A&A...565A.128C}, undetected by the \textsc{CIGALE} SED fitting.

\section{Evolution of small a-C(:H)}\label{sec:dust_evol}

\begin{figure}[t]
\centering
\includegraphics[width=9cm]{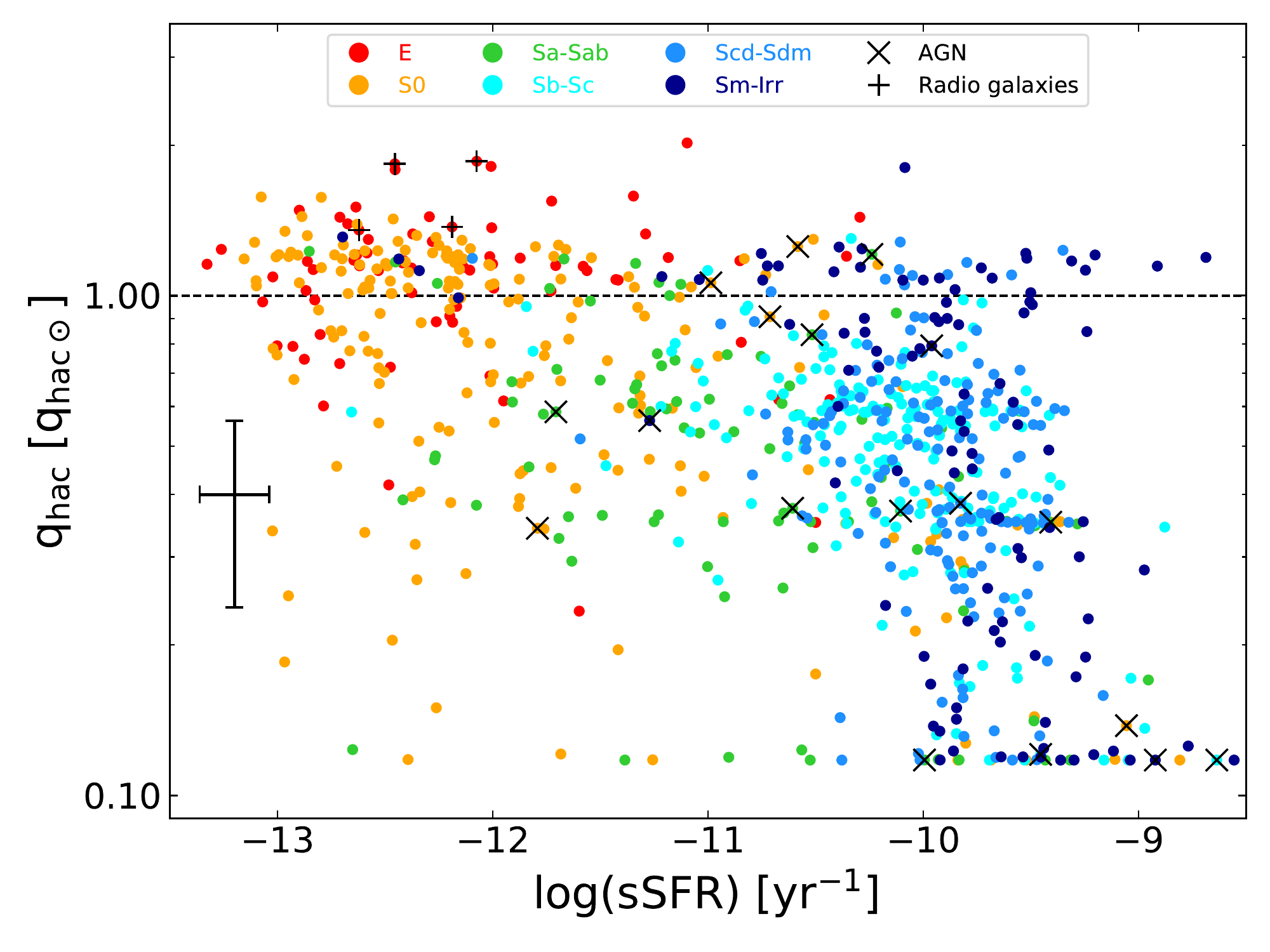}
\caption{The small a-C(:H) mass fraction ($q_\mathrm{hac}$) as a function of specific star-formation rate. The parameter $q_\mathrm{hac}$ is expressed in units of $q_\mathrm{hac\odot}$, $q_\mathrm{hac\odot}=17\%.$ Each point represents a galaxy and it is colour-coded according to morphology type, while galaxies hosting AGNs and strong radio-jets are indicated with an `X' and a `+' respectively. The large cross is the typical uncertainty on the data.}
\label{fig:qhac}
\end{figure}

Dust forms and grows in the circumstellar shells of evolved stars, e.g., asymptotic giant branch (AGB) and red giant stars or in the ejecta of core-collapse supernovae \citep{2018ARA&A..56..673G}, whilst acting as catalyst for molecular hydrogen (H$_2$) formation. On the other hand, the smallest grains, which are primarily carbonaceous nanoparticles are much more susceptible to the local conditions. This can result in their (photo/thermal) processing and possibly their complete destruction if the local physical conditions (gas temperature and density, radiation field hardness and intensity) are extreme enough \citep[see][for a review]{2004ASPC..309..347J}. Understanding the link between the dust properties and star formation, can provide useful information for galaxy evolution studies. From \textsc{CIGALE}, we were able to obtain an estimate of the mass fraction of small a-C(:H) grains, $q_\mathrm{hac}$, for our galaxy sample. We note, however, that since this parameter may not be well constrained in some cases (see the mock analysis in Appendix~\ref{ap:themis}), interpretation of the results should be considered with caution. Figure~\ref{fig:qhac} shows how $q_\mathrm{hac}$ varies with the sSFR and morphological type. The $q_\mathrm{hac}$ mass fraction is normalized to the fraction estimated for the diffuse Galactic ISM \citep[$\sim17\%$,][]{2011A&A...525A.103C}, and varies by one order of magnitude, with a typical error of 16\%. Despite the large scatter, we find a mild correlation between $q_\mathrm{hac}$ and sSFR ($\rho = -0.57$). Galaxies with low sSFR up to $\log(\text{sSFR})\sim -10.5$ have roughly constant $q_\mathrm{hac}$ values, with fractions similar or slightly higher than the one estimated for the Galactic ISM. Then, $q_\mathrm{hac}$ drops very fast for galaxies with high sSFR, $\log(\text{sSFR}) > -10.5$. \citet{2015A&A...582A.121R} found a similar decreasing trend between the PAH abundance and the sSFR for two samples of late-type galaxies (109 in total), DGS \citep[Dwarf Galaxy Survey,][]{2013PASP..125..600M} and KINGFISH.

\begin{table*}[t]
\caption{Mean luminosity ratios of the stellar and dust components of galaxies with different Hubble stages ($T$). Given that $L_\mathrm{bolo}$ is the bolometric luminosity, $L_\mathrm{dust}$ the dust luminosity,
$L_\mathrm{old}^\mathrm{unatt}$ and $L_\mathrm{young}^\mathrm{unatt}$ the unattenuated luminosity of the old and the young stars,
$L_\mathrm{old}^\mathrm{att}$ and $L_\mathrm{young}^\mathrm{att}$ the attenuated luminosity of the old and the young stars,
$L_\mathrm{old}^\mathrm{abs}$ and $L_\mathrm{young}^\mathrm{abs}$ the luminosity of the old and the young stars absorbed by dust, we define the following fractions: $f_\mathrm{old}^\mathrm{unatt}=\frac{L_\mathrm{old}^\mathrm{unatt}}{L_\mathrm{bolo}}$,
$f_\mathrm{young}^\mathrm{unatt}=\frac{L_\mathrm{young}^\mathrm{unatt}}{L_\mathrm{bolo}}$,
$f_\mathrm{old}^\mathrm{att}=\frac{L_\mathrm{old}^\mathrm{att}}{L_\mathrm{bolo}}$,
$f_\mathrm{young}^\mathrm{att}=\frac{L_\mathrm{young}^\mathrm{att}}{L_\mathrm{bolo}}$,
$f_\mathrm{abs}=\frac{L_\mathrm{dust}}{L_\mathrm{bolo}}$,
$F_\mathrm{old}^\mathrm{att}=\frac{L_\mathrm{old}^\mathrm{att}}{L_\mathrm{old}^\mathrm{unatt}}$,
$F_\mathrm{old}^\mathrm{abs}=\frac{L_\mathrm{old}^\mathrm{abs}}{L_\mathrm{old}^\mathrm{unatt}}$,
$F_\mathrm{young}^\mathrm{att}=\frac{L_\mathrm{young}^\mathrm{att}}{L_\mathrm{young}^\mathrm{unatt}}$,
$F_\mathrm{young}^\mathrm{abs}=\frac{L_\mathrm{young}^\mathrm{abs}}{L_\mathrm{young}^\mathrm{unatt}}$,
$S_\mathrm{old}^\mathrm{abs}=\frac{L_\mathrm{old}^\mathrm{abs}}{L_\mathrm{dust}}$,
and
$S_\mathrm{young}^\mathrm{abs}=\frac{L_\mathrm{young}^\mathrm{abs}}{L_\mathrm{dust}}.$
}
\begin{center}
\scalebox{1.0}{
\begin{tabular}{c|cc|ccc||cc|cc||cc}
\hline 
\hline 
$T$ &  
$f_\mathrm{old}^\mathrm{unatt}$ & $f_\mathrm{young}^\mathrm{unatt}$ & $f_\mathrm{old}^\mathrm{att}$ & $f_\mathrm{young}^\mathrm{att}$ & $f_\mathrm{abs}$ & 
$F_\mathrm{old}^\mathrm{att}$ & $F_\mathrm{old}^\mathrm{abs}$ &  $F_\mathrm{young}^\mathrm{att}$ & $F_\mathrm{young}^\mathrm{abs}$ &
$S_\mathrm{old}^\mathrm{abs}$ & $S_\mathrm{young}^\mathrm{abs}$ \\
\hline
  -5 & 0.99 & 0.01 & 0.97 & 0.01 & 0.02 & 0.98 & 0.02 & 0.84 & 0.16 & 0.91 & 0.09\\
  -4 & 0.98 & 0.02 & 0.97 & 0.01 & 0.02 & 0.99 & 0.01 & 0.87 & 0.13 & 0.90 & 0.10\\
  -3 & 0.93 & 0.07 & 0.88 & 0.03 & 0.09 & 0.93 & 0.07 & 0.66 & 0.34 & 0.84 & 0.16\\
  -2 & 0.96 & 0.04 & 0.92 & 0.02 & 0.06 & 0.95 & 0.05 & 0.73 & 0.27 & 0.88 & 0.12\\
  -1 & 0.94 & 0.06 & 0.86 & 0.02 & 0.12 & 0.90 & 0.10 & 0.55 & 0.45 & 0.84 & 0.16\\
   0 & 0.94 & 0.06 & 0.86 & 0.02 & 0.12 & 0.91 & 0.09 & 0.52 & 0.48 & 0.79 & 0.21\\
   1 & 0.89 & 0.11 & 0.73 & 0.02 & 0.25 & 0.79 & 0.21 & 0.31 & 0.69 & 0.76 & 0.24\\
   2 & 0.87 & 0.13 & 0.72 & 0.03 & 0.25 & 0.80 & 0.20 & 0.30 & 0.70 & 0.69 & 0.31\\
   3 & 0.81 & 0.19 & 0.62 & 0.05 & 0.33 & 0.76 & 0.24 & 0.23 & 0.77 & 0.59 & 0.41\\
   4 & 0.79 & 0.21 & 0.63 & 0.06 & 0.31 & 0.79 & 0.21 & 0.28 & 0.72 & 0.54 & 0.46\\
   5 & 0.75 & 0.25 & 0.59 & 0.07 & 0.34 & 0.78 & 0.22 & 0.28 & 0.72 & 0.48 & 0.52\\
   6 & 0.77 & 0.23 & 0.65 & 0.11 & 0.24 & 0.85 & 0.15 & 0.45 & 0.55 & 0.47 & 0.53\\
   7 & 0.75 & 0.25 & 0.64 & 0.11 & 0.25 & 0.85 & 0.15 & 0.42 & 0.58 & 0.44 & 0.56\\
   8 & 0.77 & 0.23 & 0.69 & 0.13 & 0.18 & 0.89 & 0.11 & 0.55 & 0.45 & 0.44 & 0.56\\
   9 & 0.73 & 0.27 & 0.67 & 0.18 & 0.15 & 0.91 & 0.09 & 0.67 & 0.33 & 0.40 & 0.60\\
  10 & 0.73 & 0.27 & 0.68 & 0.20 & 0.12 & 0.94 & 0.06 & 0.74 & 0.26 & 0.44 & 0.56\\
\hline
    $\left[-5.0,-3.5\right)$ & 0.98 & 0.02 & 0.97 & 0.01 & 0.02 & 0.98 & 0.02 & 0.85 & 0.15 & 0.90 & 0.10 \\
    $\left[-3.5,0.5\right)$  & 0.95 & 0.05 & 0.89 & 0.02 & 0.09 & 0.93 & 0.07 & 0.64 & 0.36 & 0.85 & 0.15 \\
    $\left[0.5,2.5\right)$   & 0.88 & 0.12 & 0.72 & 0.03 & 0.25 & 0.80 & 0.20 & 0.31 & 0.69 & 0.73 & 0.27 \\
    $\left[2.5,5.5\right)$   & 0.78 & 0.22 & 0.61 & 0.06 & 0.33 & 0.78 & 0.22 & 0.27 & 0.73 & 0.53 & 0.47 \\
    $\left[5.5,8.5\right)$   & 0.77 & 0.23 & 0.66 & 0.11 & 0.23 & 0.86 & 0.14 & 0.46 & 0.54 & 0.45 & 0.55 \\
    $\left[8.5,10.0\right]$  & 0.73 & 0.27 & 0.68 & 0.19 & 0.13 & 0.93 & 0.07 & 0.72 & 0.28 & 0.42 & 0.58 \\
\hline \hline
\end{tabular}}
\label{tab:old_young_dust}
\end{center}
\end{table*}

The behaviour seen in Fig.~\ref{fig:qhac} indicates that galaxies at the beginning of their evolutionary stage have low mass fractions of small grains and as galaxies evolve, more metals become available in the ISM resulting to a more efficient grain growth, and thus to higher $q_\mathrm{hac}$ fractions. The majority of low mass LTGs (Sm-Irr) in Fig.~\ref{fig:qhac}, have on average higher sSFRs (see Table~\ref{tab:dsr}), and significantly lower $q_\mathrm{hac}$ values. In this case, galaxies with high sSFR possess stronger UV radiation fields, often resulting to the efficient destruction of the small grains in the ISM. As `typical' spiral galaxies grow in stellar mass and gas, their a-C(:H) mass fractions increase too, with $q_\mathrm{hac}$ reaching up to $q_\mathrm{hac\odot}$, and then remain roughly constant for the later evolutionary stages. This behaviour indicates a balance between dust destruction and dust growth \citep{2014MNRAS.440.1562M}, consistent with the findings of \citet{2019A&A...623A...5D}, which studied the variation of the dust-to-metal ratio for a subsample of $\sim 500$~DustPedia galaxies. They found that a chemical evolution model with a significant contribution from grain growth describes DustPedia galaxies fairly well, with the more evolved galaxies having a constant dust-to-metal ratio, while less evolved galaxies have on average 2.5 times lower dust-to-metal ratios.

\section{The stellar populations and the dust content in nearby galaxies} \label{sec:stardust}

A knowledge of the different components of the stellar populations in galaxies and the way their released energy is interacting with the dust particles is a crucial piece of information in understanding the full picture of galaxy formation and evolution \citep{2012ApJ...748..123S, 2014A&A...571A..72B, 2015MNRAS.448..135B}. Here we explore the importance of the old and the young stellar components of the galaxies as parametrized by \textsc{CIGALE}. \textsc{CIGALE} distinguishes between two classes of stars, one of a variable age accounting for the average old stellar population and one of a younger age ($t_\mathrm{flex}$ as parametrized in the case of a flexible-delayed SFH used here) accounting for the young stellar population. The old stellar population was given the freedom to obtain values between five different ages ranging from 2 to 12~Gyr (see Table~\ref{tab:param}, $t_\mathrm{gal}$) while the age of 200~Myr was fixed for the young stellar population. The choice of a fixed young stellar age was dictated by the need to keep the total number of parameters in \textsc{CIGALE} to a minimum and from the fact that varying the age will not significantly alter the shape of the SED and thus the luminosity of this component \citep[see][and their Fig.~3 where SEDs of a range of $t_\mathrm{trunc}$ values are plotted]{2016A&A...585A..43C}.

\begin{figure*}
\centering
\includegraphics[width=18cm]{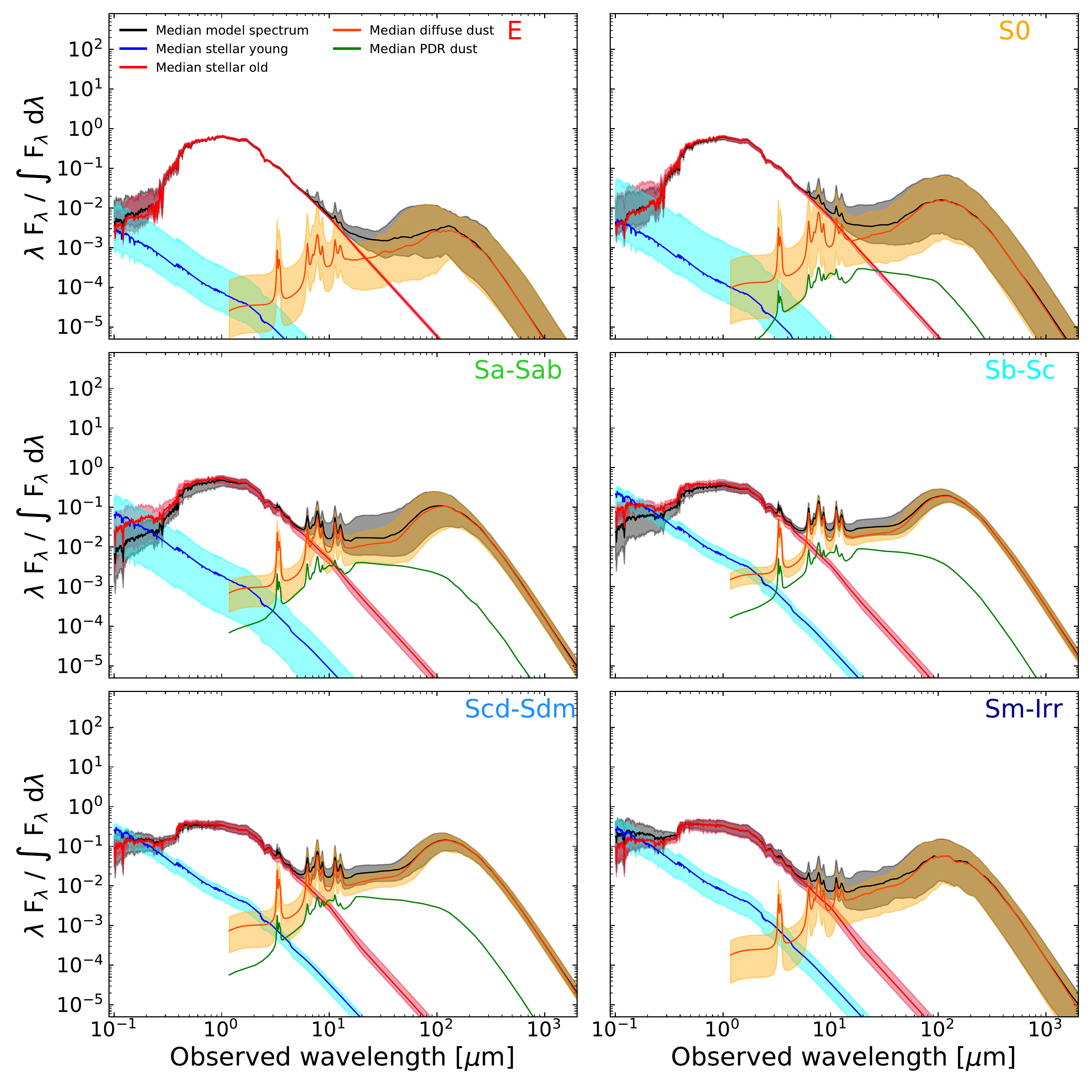}
\caption{Template SEDs for the six main morphological classes as derived by \textsc{CIGALE}. The median SED of each sub-class is shown as a black curve while the unattenuated SEDs of the old and the young stars are shown as red and blue curves respectively. The orange curve indicates the median spectrum of the diffuse dust, while the green curve shows the emission from PDRs. The shaded areas represent the range of the 16-84th percentiles to the median value (except for the case of the PDR spectra, where, for clarity, we refrain from presenting the full range of SEDs). For each subsample, the $10\%$ most deviant SEDs have been excluded \citep[see also the template SEDs presented in][]{2018A&A...620A.112B}.}
\label{fig:cigale_seds}
\end{figure*}

In Fig.~\ref{fig:cigale_seds} we present the median SEDs fitted by \textsc{CIGALE} for six main morphological types. In each panel the median SED for each Hubble type is indicated as a black solid curve. The red and blue curves in the optical part of the SED are the median SEDs of the unattenuated old and young stellar populations respectively, while the orange and green curves in the MIR-submm part are the median SEDs indicating the diffuse dust and the dust in photo-dissociation regions respectively. In all curves the shaded areas bracket the 16th and 84th percentiles around the median. A first visual inspection of the SEDs of different Hubble types shows that the young stellar population becomes less dominant, as compared to the old stellar population, when following the evolutionary track from the late-type and irregular galaxies to the early-type galaxies. This can be quantified by calculating the relative contribution of each component to the total bolometric luminosity of each galaxy, something that we will discuss in Sect.~\ref{sec:old_young}. 

In the FIR-submm part of the spectrum we see that, for all types of galaxies, the emission is dominated by the diffuse dust component (orange curve) with the PDR dust (green curve) being only a small fraction mainly contributing to the MIR emission. In the cases of E and Sm-Irr type galaxies, in particular, the PDR dust emission is negligible with the MIR emission composed mainly from the superposition of the diffuse dust emission and the emission originating from the old stellar population. The diffuse dust emission progressively becomes a significant part of the bolometric luminosity of the galaxy when moving along the evolutionary track from E to Sb-Sc but then it becomes less prominent at later type galaxies. As a result of the large quantities of dust grain material in these galaxies a severe extinction is observed (especially evident in the UV wavelengths, below $\sim0.4~\mu$m). This is a striking feature in the SEDs of Sa-Sab, and Sb-Sc galaxies where the unattenuated SEDs of both the old and the young components (red and blue respectively) are exceeding the `observed' SED of the galaxy (black curve). In E and S0 galaxies the extinction is minimal with both unattenuated curves below the observed SED, while for Scd-Sdm and Sm-Irr galaxies it is only the young stars that suffer from significant attenuation at UV wavelengths below $\sim0.2~\mu$m.

\subsection{Old and young stellar populations in galaxies} \label{sec:old_young}

\begin{figure*}
\centering
\includegraphics[width=18cm]{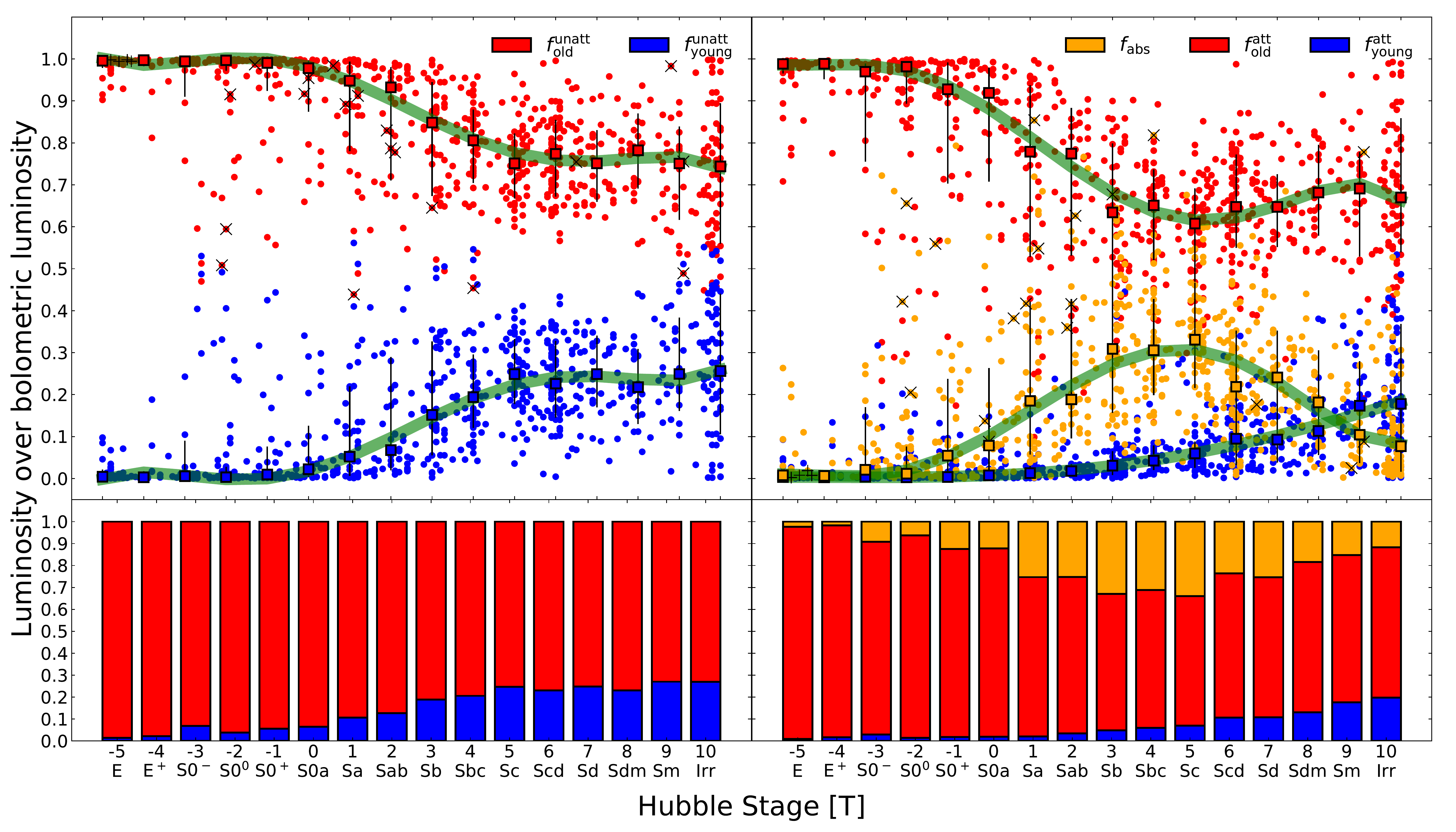}
\caption{Left: The ratio of the unattenuated luminosity of the old and the young stellar components to the bolometric luminosity (red and blue circles respectively; top-left panel) along with the respective stacked-bar plots of the mean values (bottom-left panel). Right: The ratio of the attenuated luminosity of the old and the young stellar components to the bolometric luminosity (red and blue circles respectively) together with the ratio of the dust luminosity to the bolometric luminosity (orange circles; top-right panel) along with the respective stacked-bar plots of the mean values (bottom-right panel). Square symbols in the top panels are the median values for a specific Hubble stage bin while the bars indicate the 16th and 84th percentiles range. The thick green curves are the 5th order polynomial regression to the median values (see Table~\ref{tab:spline} for the polynomial regression parameters). Galaxies hosting an AGN or strong radio-jets are marked with an `X' or a `+', respectively, for the old stellar population only (red circles) in the top-left panel, and for the dust luminosity only (orange circles) in the top-right panel.}
\label{fig:old_young_dust}
\end{figure*}

Investigating the relative contribution of the old and the young stellar components to the bolometric luminosity for galaxies of different morphological types as well as the effect of the different stellar populations in the dust heating is a difficult task. This is because the light originating directly from the stars cannot directly be observed due to its attenuation by the dust. The only way to overcome this problem is by exploiting the information hidden in the SEDs using appropriate methods that simultaneously treat the stellar and dust emission. We do this based on the parametrization of the stellar populations and the dust emission obtained by fitting \textsc{CIGALE} to the DustPedia galaxies. This is shown in the top-left panel of Fig.~\ref{fig:old_young_dust} with the ratio of the unattenuated luminosities (i.e., the intrinsic luminosities) of the two stellar components (old and young) to the bolometric luminosity ($f_\mathrm{old}^\mathrm{unatt}=L_\mathrm{old}^\mathrm{unatt}/L_\mathrm{bolo}$, $f_\mathrm{young}^\mathrm{unatt}=L_\mathrm{young}^\mathrm{unatt}/L_\mathrm{bolo}$) plotted as red and blue circles respectively. What is immediately evident is the dominant role of the old, more evolved, stars in the total luminosity. Independent from the morphology of the galaxy, on average, the old stellar component dominates the bolometric luminosity of the galaxy contributing with more than $\sim75\%$. The luminosity of the ETGs ($T<0.5$) is the most extreme example, dominated by the emission of the old stars and with only a small contribution (maximum of $\sim10\%$ at $T=0$) from the young stars (see Table~\ref{tab:old_young_dust}). For a detailed discussion on the contribution of old and young stars on the luminosity of ETGs (focusing on the different behavior of the two `classical' subsamples of elliptical and lenticular galaxies) we refer to Cassar\`{a} et al. (2019, to be submitted). For morphological types from $T=0-5$, there is a gradual rise in the contribution of the young stars to the bolometric luminosity reaching about $25\%$ while it stays practically constant for morphological types $T>5$. The scatter in each morphological bin is quite significant, though a clear trend, as discussed above, is visible. 

Some galaxies, of various Hubble stages, show values of the relative luminosities of the different stellar components to the bolometric luminosities, reaching up to $50\%$, for example NGC~1222, NGC~1377, ESO097-013, ESO~493-016, NGC~2993, NGC~4194, NGC~6300, and NGC~7714. These are all galaxies showing extreme values of SFR, well above the average value for each morphological bin. The bottom-left panel in Fig.~\ref{fig:old_young_dust} shows the stacked bars of the mean values per Hubble stage with the exact mean values provided in Table~\ref{tab:old_young_dust} (columns $f_\mathrm{old}^\mathrm{unatt}$ and $f_\mathrm{young}^\mathrm{unatt}$).

\subsection{The heating of dust by the different stellar populations}

\begin{figure*}[t]
\centering
\includegraphics[width=18cm]{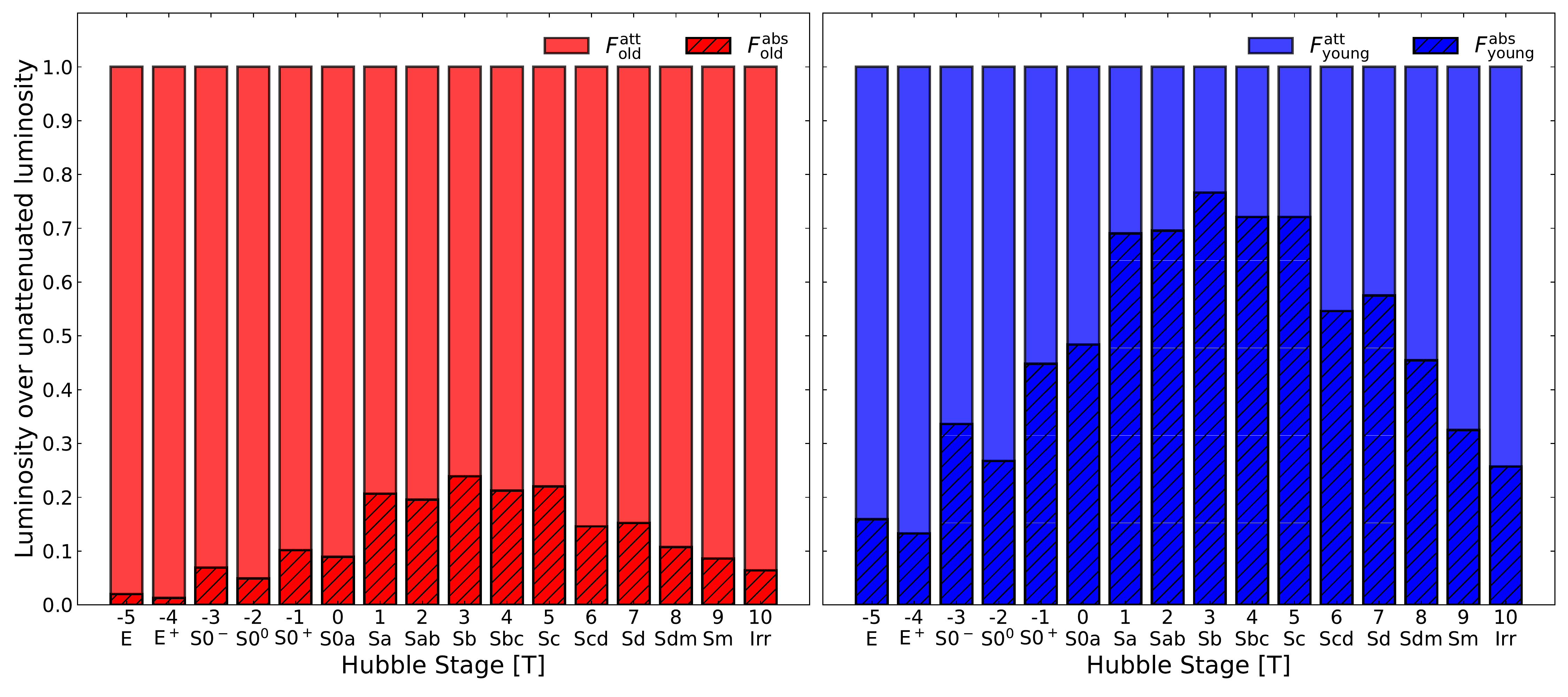}
\caption{Mean values, per Hubble stage bin, of the fraction of the luminosity of the old and the young stellar components (left and right panels respectively) used for the dust heating. In each panel the mean values of the ratio of the luminosity absorbed, by the dust, to the unattenuated luminosity of the specific stellar component ($F_\mathrm{old}^\mathrm{abs}$, $F_\mathrm{young}^\mathrm{abs}$) is shown as crossed bars, while the mean values of the ratio of the attenuated luminosity of the specific stellar component to the unattenuated luminosity 
($F_\mathrm{old}^\mathrm{att}$, $F_\mathrm{young}^\mathrm{att}$) is shown as solid bars. The mean values are provided in Table~\ref{tab:old_young_dust}.}
\label{fig:old_young_att_abs}
\end{figure*}

The presence of large quantities of dust material throughout the galaxy affects how the galaxy is observed, by extinguishing the light originating from the different stellar populations. The lost energy will be deposited into the dust giving rise to the luminosity at FIR wavelengths. This interplay between stars and dust is presented in the top-right panel of Fig.~\ref{fig:old_young_dust}, in its simplest way, for galaxies of different Hubble stages. The red and blue circles now indicate the luminosity attenuated by dust, normalized to the bolometric luminosity of each galaxy, for the old and the young stars respectively ($f_\mathrm{old}^\mathrm{att}=L_\mathrm{old}^\mathrm{att}/L_\mathrm{bolo}$, $f_\mathrm{young}^\mathrm{att}=L_\mathrm{young}^\mathrm{att}/L_\mathrm{bolo}$). Comparing with the top-left panel of Fig.~\ref{fig:old_young_dust}, a decrease in luminosity is observed for both stellar populations, which is most important in intermediate Hubble stages ($1\le T \le7$). This energy is absorbed by the dust and re-emitted in the IR and submm wavelengths giving rise to the dust luminosity (orange circles; $f_\mathrm{abs}=L_\mathrm{dust}/L_\mathrm{bolo}$). 

The bottom-right panel in Fig.~\ref{fig:old_young_dust} shows the stacked bars of the mean value per Hubble stage for the three components (old stars, young stars and dust) with the exact mean values provided in Table~\ref{tab:old_young_dust} (columns $f_\mathrm{old}^\mathrm{att}$, $f_\mathrm{young}^\mathrm{att}$ and $f_\mathrm{abs}$). What is interesting to notice from this plot is the continuous, monotonic increase of $f_\mathrm{young}^\mathrm{att}$ when following the Hubble sequence from E to Irr galaxies reaching maximum mean values of $20\%$ for Irr galaxies. Both the young and the old stars are mostly affected by dust in intermediate spiral galaxies ($1\le T \le7$) with a drop of more than $15\%$ in their intrinsic luminosities.  

To further investigate the efficiency of the two stellar populations in the heating of the dust, we calculated the ratios of the attenuated and the absorbed, by the dust, luminosities in each stellar component to the unattenuated luminosity of the specific component ($F_\mathrm{old}^\mathrm{att}=L_\mathrm{old}^\mathrm{att}/L_\mathrm{old}^\mathrm{unatt}$, $F_\mathrm{old}^\mathrm{abs}=L_\mathrm{old}^\mathrm{abs}/L_\mathrm{old}^\mathrm{unatt}$ for the old stellar population and $F_\mathrm{young}^\mathrm{att}=L_\mathrm{young}^\mathrm{att}/L_\mathrm{young}^\mathrm{unatt}$ and $F_\mathrm{young}^\mathrm{abs}=L_\mathrm{young}^\mathrm{abs}/L_\mathrm{young}^\mathrm{unatt}$ for the young stellar population). This is presented in the stacked bar graphs of Fig.~\ref{fig:old_young_att_abs}, with the left panel for the relative contribution of the old stars and the right panel for the young stars. In each panel the shaded bars show the fraction of the luminosity of each stellar component that is absorbed by the dust and contribute to its heating, while the remaining is the fraction of the luminosity that is left as direct light emitted by the stars. For the case of the old stellar population the stars donate up to $\sim24\%$ of their luminosity for Sb galaxies to the dust heating, with a very small fraction of their luminosity (below $\sim10\%$) being used for this purpose for the two extremes of the Hubble stages (E and Irr). The young stars, on the other hand, are more generous giving a significant fraction of their luminosity in heating the dust which can be up to $\sim77\%$, again for Sb galaxies, with more than $\sim15\% - 20\%$ for the extreme Hubble stages.     

\begin{figure}[ht]
\centering
\includegraphics[width=9cm]{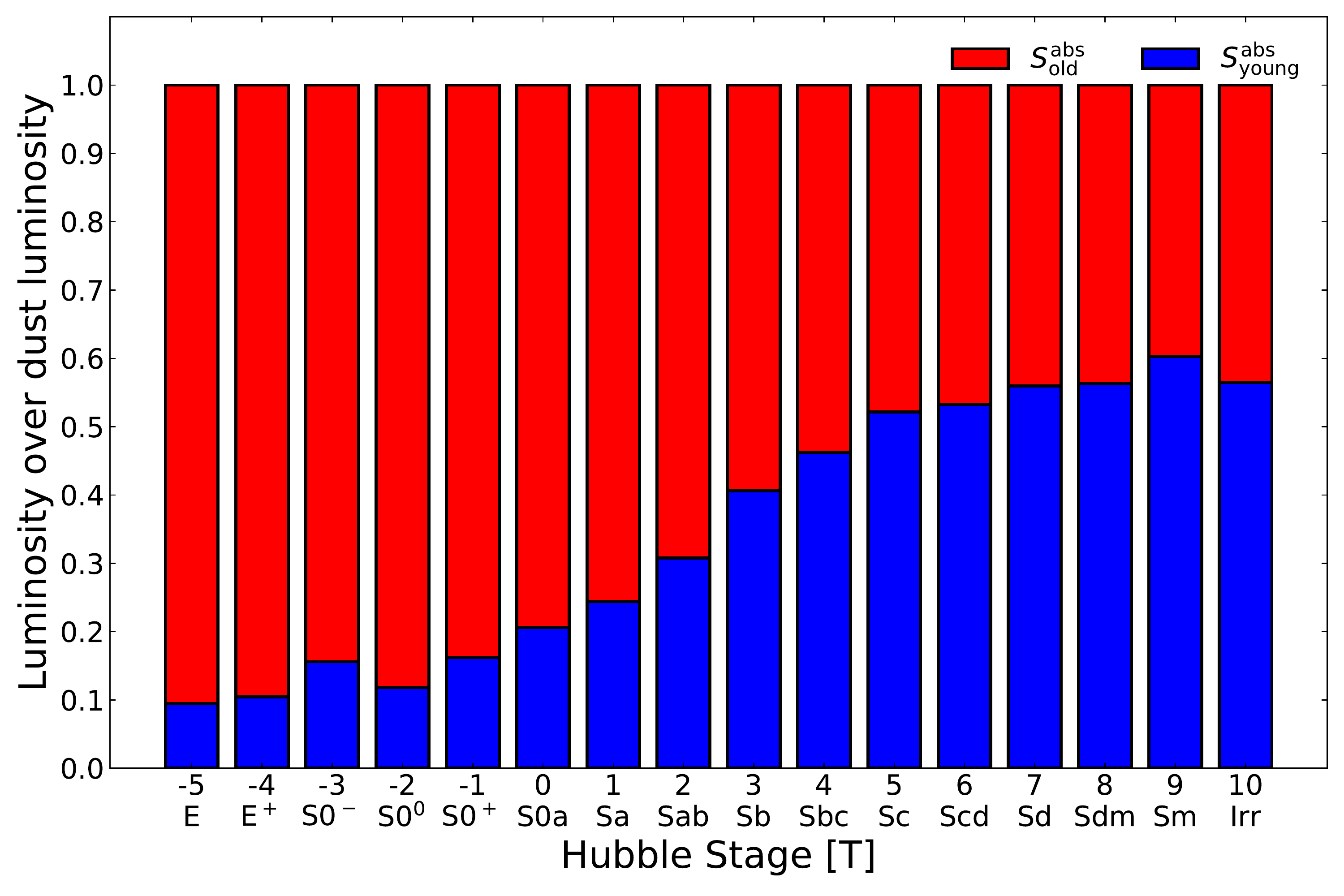}
\caption{The mean values per Hubble stage bin, of the ratios of the luminosity absorbed by dust, for a specific stellar component to the dust luminosity. Red and blue bars refer to the old and the young stellar components ($S_\mathrm{old}^\mathrm{abs}$ and $S_\mathrm{young}^\mathrm{abs}$) respectively. The mean values are provided in Table~\ref{tab:old_young_dust}.}
\label{fig:old_young_abs}
\end{figure}

The relative contribution of the old and the young stars to the dust heating is shown in Fig~\ref{fig:old_young_abs} with the absorbed luminosities to the dust luminosity ($S_\mathrm{old}^\mathrm{abs}=L_\mathrm{old}^\mathrm{abs}/L_\mathrm{dust}$, and $S_\mathrm{young}^\mathrm{abs}=L_\mathrm{young}^\mathrm{abs}/L_\mathrm{dust}$ for the old and the young stellar populations respectively) plotted as stacked bars (red and blue respectively). Here, it is interesting to notice the gradual increase in the contribution of the young stars to the dust heating from only $\sim10\%$ for E galaxies to $\sim60\%$ for later type galaxies (see Table~\ref{tab:old_young_dust} for the exact values). To obtain a clearer view on the dust heating processes, we look at the relation between the $S_\mathrm{young}^\mathrm{abs}$ and the sSFR. The sSFR is sensitive to the different heating sources in galaxies and can adequately trace the hardness of the UV radiation field \citep{2014A&A...565A.128C}.

\begin{figure}[t]
\centering
\includegraphics[width=9cm]{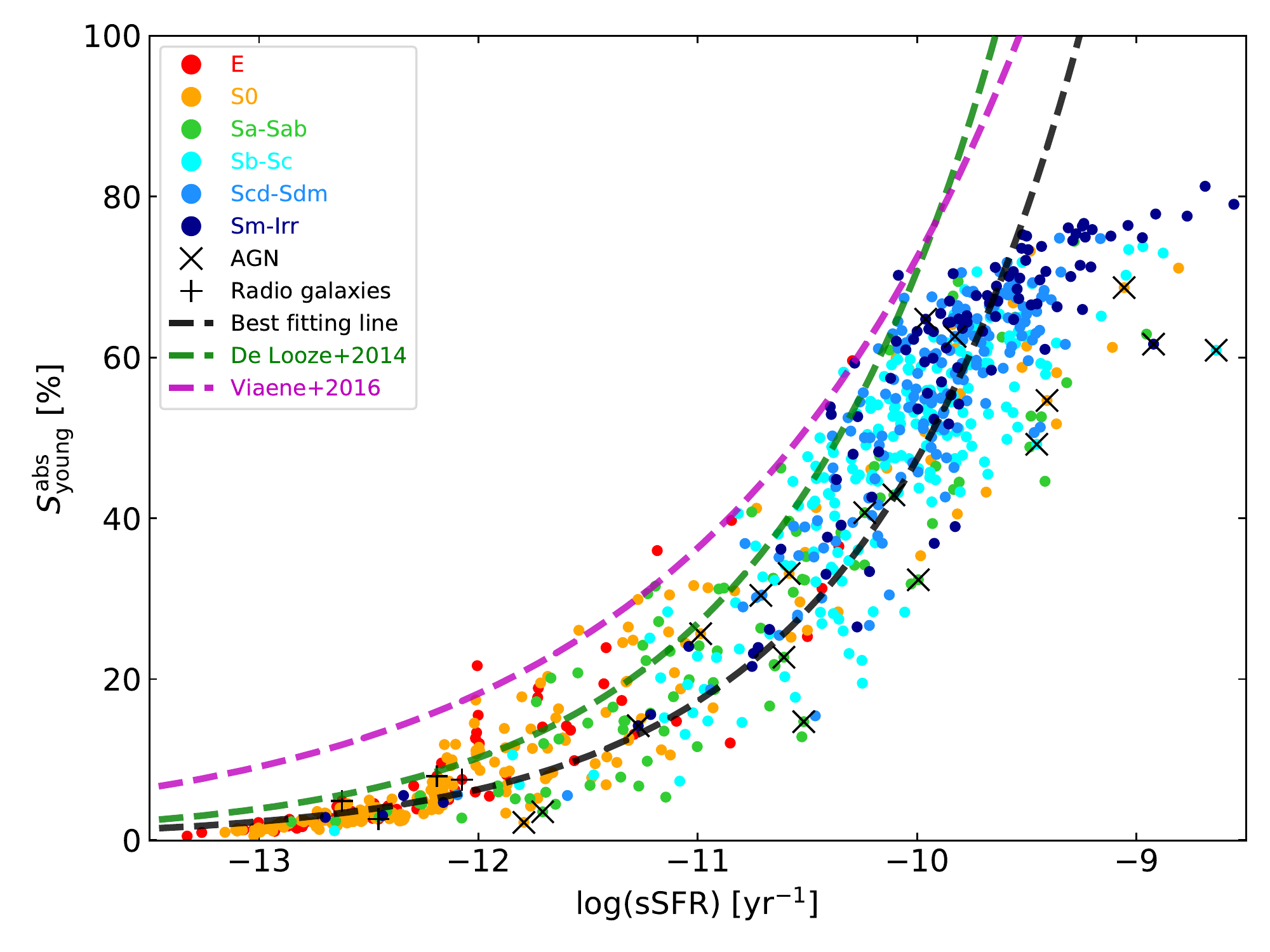}
\caption{Correlation between $S_\mathrm{young}^\mathrm{abs}$ and sSFR. Each point represents a galaxy and it is colour-coded according to morhphological type. Galaxies hosting AGNs and strong radio-jets are marked with an `X' and a `+' respectively. The black dashed line is the best fitted powerlaw. For comparison, the best fitted lines for M51 \citep[green line:][]{2014A&A...571A..69D} and HRS \citep[magenta line:][]{2016A&A...586A..13V} are shown.}
\label{fig:funev}
\end{figure}

Figure~\ref{fig:funev} shows the correlation between the sSFR and $S_\mathrm{young}^\mathrm{abs}$. We find a tight correlation with a Spearman’s coefficient $\rho = 0.95$. The relation between the two quantities can be approximated with a power law function:

\begin{equation} \label{eq:powerlaw}
\log S_\mathrm{young}^\mathrm{abs} = 0.44 \times \log sSFR + 6.06 \, ,
\end{equation}

Galaxies with $\log \text{sSFR}>-10.5$ have high $S_\mathrm{young}^\mathrm{abs}$ fractions, indicating that dust is mainly heated by UV radiation emitted by the young stellar population, whereas galaxies with low $\log \text{sSFR}$ have extremely low heating fractions. A similar correlation was reported by \citet{2014A&A...571A..69D} and \citet{2017A&A...599A..64V}. They used radiation transfer simulations to quantify the dust heating fraction due to the young stellar population ($\le 100$~Myr) for M51 and the Andromeda galaxy respectively. They found that high heating fractions correspond to high levels of sSFR. Furthermore, \citet{2016A&A...586A..13V} also reported a similar trend between the UV dust heating fraction and the sSFR for 239 LTGs of the HRS sample. A clear offset can be seen in Fig.~\ref{fig:funev} between the results of this work (black line) and the work of \citet{2014A&A...571A..69D} (green line) and \citet{2016A&A...586A..13V} (magenta line). We attribute this offset due to the different methods used to estimate the sSFR and the heating fraction. \citet{2014A&A...571A..69D} estimated the star-formation rate of M51 in a pixel-by-pixel basis, using the MAPPINGS \citep{2004ApJS..153....9G} SED templates of the ionizing stars ($\le 10~$Myr), while \citet{2016A&A...586A..13V} used \textsc{MAGPHYS} to derive the sSFR. Furthermore, we also notice a correlation of the heating fraction with morphology, with LTGs having the highest heating fractions. From this analysis, it is evident that dust heating is driven globally and locally, by the ratio of ongoing star-formation, and the past star-formation.

The approach we used in this study to determine the dust heating due to the different stellar populations is only a rough approximation given the perplexing interaction among the stellar radiation field and the dust. Other sources, not treated here, could potentially contribute to the dust heating [e.g, the X-ray emission from a hot halo gas \citep{2010ApJ...725..955N} or an AGN]. Thus, a more sophisticated approach is needed to describe the complicated processes that contribute to the heating of dust in galaxies that take into account the effect of non-local heating. Such approaches include detailed treatment of the 3D morphology of galaxies through radiative transfer modelling. Following previous studies by \citet{2014A&A...571A..69D} and \citet{2017A&A...599A..64V}, several DustPedia galaxies have been modelled using the \textsc{SKIRT} radiative transfer code \citep{2011ApJS..196...22B, 2015A&C.....9...20C}. These galaxies include the most extended, face-on galaxies in our sample [e.g., NGC~3031 (M~81), NGC~1365, NGC~5236 (M~83), NGC~3351 (M~95), NGC~4321 (M~100) and NGC~1068 (M~77)] with a variety of morphologies (grand design spirals, lenticulars, barred, AGNs). Although the heating of the dust (attributed to old and young stars) in these studies is investigated on local scales, our results on global scales agree fairly well [details of this analysis are presented in Verstocken et al. (to be submitted), Nersesian et al. (in preparation) and Viaene et al. (in preparation)]. 

\section{Summary and conclusions} \label{sec:sum_con}

We have used photometric measurements of a sample of 814 local galaxies, drawn from the DustPedia archive, to construct their multi-wavelength SEDs (from the UV to the submm). This is the first dedicated dust heating study for such a large, statistically significant sample. The majority of these objects ($94\%$) have more than 15 such measurements available (with a minimum of 10 and a maximum of 30). The galaxies span a variety of morphologies parametrized with their Hubble stage ($T$) on a scale from -5 to 10 (from pure ellipticals to irregular galaxies respectively) with an average of 50 objects in each morphology class. In order to extract information on their baryonic content (stars and dust) we utilize the advanced fitting tool \textsc{CIGALE} adapted so as to include the recently developed dust model \textsc{THEMIS} \citep{2017A&A...602A..46J} that successfully explains the observed FUV-NIR extinction, the IR to submm dust thermal emission and the shape of the infrared emission bands. For each galaxy we obtain accurate measurements of the stellar mass, the current star-formation rate, the dust mass, and the dust temperature, while the stellar populations in each galaxy and their role in the dust heating is investigated by deriving the luminosity produced by the old and the young stars separately. Additional information on the atomic gas mass ($M_\text{HI}$) for a subsample of 711 galaxies is also provided for each galaxy \citep{2019A&A...623A...5D}. For comparison, we have derived the global dust properties (mass and temperature) in an independent way, by fitting the FIR-submm part of the SED ($\lambda \ge 100~\mu$m) with a modified black-body properly scaled to account for the \textsc{THEMIS} dust physics. Our analysis indicates that:

\begin{itemize}

\item The stellar mass is maximal for pure ellipticals ($T=-5$) and small variations for galaxies of $T<2$ with a sharp drop (of about two orders of magnitude) for later-type galaxies. The atomic gas mass varies slightly for galaxies with $T<2$ (very similar to the stellar mass) followed by a drop (of about an order of a magnitude) for later-type galaxies. The dust mass and SFR change in a similar way, between different morphological classes, with a continuous increase for earlier-types and a decrease for later-type galaxies (of about two orders of magnitude in both cases), with a peak value for galaxies around $T=5$. 

\item Normalization to the stellar mass of the galaxy shows an increasing trend (from $T=-5$ to $T=10$) for both the dust and the gas content as well as the SFR with $M_\mathrm{dust}^\textsc{CIGALE}$/$M_\mathrm{star}$ obtaining its maximum value at $T=7$, $M_\mathrm{HI}$/$M_\mathrm{star}$ increasing continuously from $T=-5$ to $T=10$ and with sSFR being roughly constant for ETGs with $T<-2$, increasing rapidly for galaxies with Hubble stages up to $T=5$ followed by a mild increase for later-type galaxies. The dust-to-gas mass ratio ($M_\mathrm{dust}^\textsc{CIGALE}$/$M_\mathrm{HI}$) obtains its maximum value at around $T=2$ with lower values (by about two orders of magnitude) for earlier- and later-type galaxies.
 
\item The dust temperature, calculated by scaling the strength of the ISRF intensity derived by \textsc{CIGALE}, is higher for ETGs ($\sim30~$K) compared to LTGs where a drop by $\sim10~$K is observed, followed by a sharp rise back to $\sim30~$K for Sm-Irr type galaxies. The dust temperatures compare fairly well with those derived by fitting a single MBB in the wavelength range 100-600~$\mu$m, especially for LTGs.

\item The mass fraction of aromatic feature emitting grains $q_\mathrm{hac}$, correlates with sSFR and morphology. High sSFR galaxies have low dust mass fractions and as galaxies grow in stellar mass and gas content, $q_\mathrm{hac}$ rises up to values close to the one estimated for the Galactic ISM ($\sim17\%$). For galaxies with sSFR~$>10^{-10}~\text{yr}^{-1}$, $q_\mathrm{hac}$ was found to be roughly constant. 

\item ETGs contain, mainly, old stars with only a small fraction of the bolometric luminosity ($<10\%$) originating from young stars. For spiral galaxies with Hubble stages from 0-5 the fraction of young stars gradually increases up to $\sim25\%$, while it stays roughly constant for galaxies with Hubble stages larger than 5. 

\item The dust luminosity normalized to the bolometric luminosity of the galaxy ($f_\mathrm{abs}$) gets its maximum value ($\sim34\%$) for galaxies with Hubble stages around 5 while it progressively gets down to nearly zero for ellipticals ($T=-5$) and to $\sim10\%$ for irregulars ($T=10$) \citep[see][for a complete review on $f_\mathrm{abs}$]{2018A&A...620A.112B}. 

\item On average, young stars are heating the dust more efficiently with the absorbed, by the dust, luminosity reaching as high as $\sim77\%$ (at $T=3$) of the total, unattenuated luminosity of the young stars. On the other hand, the maximum luminosity of the old stars used in the heating of dust is $\sim24\%$, again at $T=3$. 

\item The heating of the dust in ETGs is dominated by the old stars to a level of up to $\sim90\%$ while the young stars progressively contribute more to the dust heating for galaxies with Hubble stages from 0 to 5, while they become the dominant source of dust heating for galaxies with Hubble stages greater than 5 contributing to $\sim60\%$ to the dust heating.

\item The dust heating fraction by young stars is strongly correlated with the specific star-formation rate with higher heating fractions corresponding to higher sSFR. There is also a clear trend between the heating fraction and the morphological type with significantly higher fractions in late type galaxies.

\end{itemize}

Recipes to estimate physical properties of galaxies, as a function of the Hubble stage of the galaxy are provided in Table~\ref{tab:spline}. The results of the \textsc{CIGALE} and the MBB modelling, presented in this study, are provided in the DustPedia archive for every galaxy modelled.

\section*{Acknowledgements}

We would like to thank the anonymous referee for providing comments and suggestions that helped to improve the quality of the manuscript.

DustPedia is a collaborative focused research project supported by the European Union under the Seventh Framework Programme (2007-2013) call (proposal no. 606847). The participating institutions are: Cardiff University, UK; National Observatory of Athens, Greece; Ghent University, Belgium; Université Paris Sud, France; National Institute for Astrophysics, Italy and CEA, France. 

This research is co-financed by Greece and the European Union (European Social Fund- ESF) through the Operational Programme «Human Resources Development, Education and Lifelong Learning» in the context of the project “Strengthening Human Resources Research Potential via Doctorate Research“ (MIS-5000432), implemented by the State Scholarships Foundation (IKY). 

M. Decleir, W. Dobbels, I. De Looze and S. Viaene gratefully acknowledge the support of the Research Foundation - Flanders (FWO Vlaanderen). 

We would like to thank M. Boquien and L. Ciesla for providing useful information and valuable help in using \textsc{CIGALE}.

\bibliographystyle{aa}
\bibliography{main.bbl}

\appendix
%

\newpage
\section{Mock Analysis}\label{ap:themis}

\begin{figure}[!h]
\centering
\includegraphics[width=9cm]{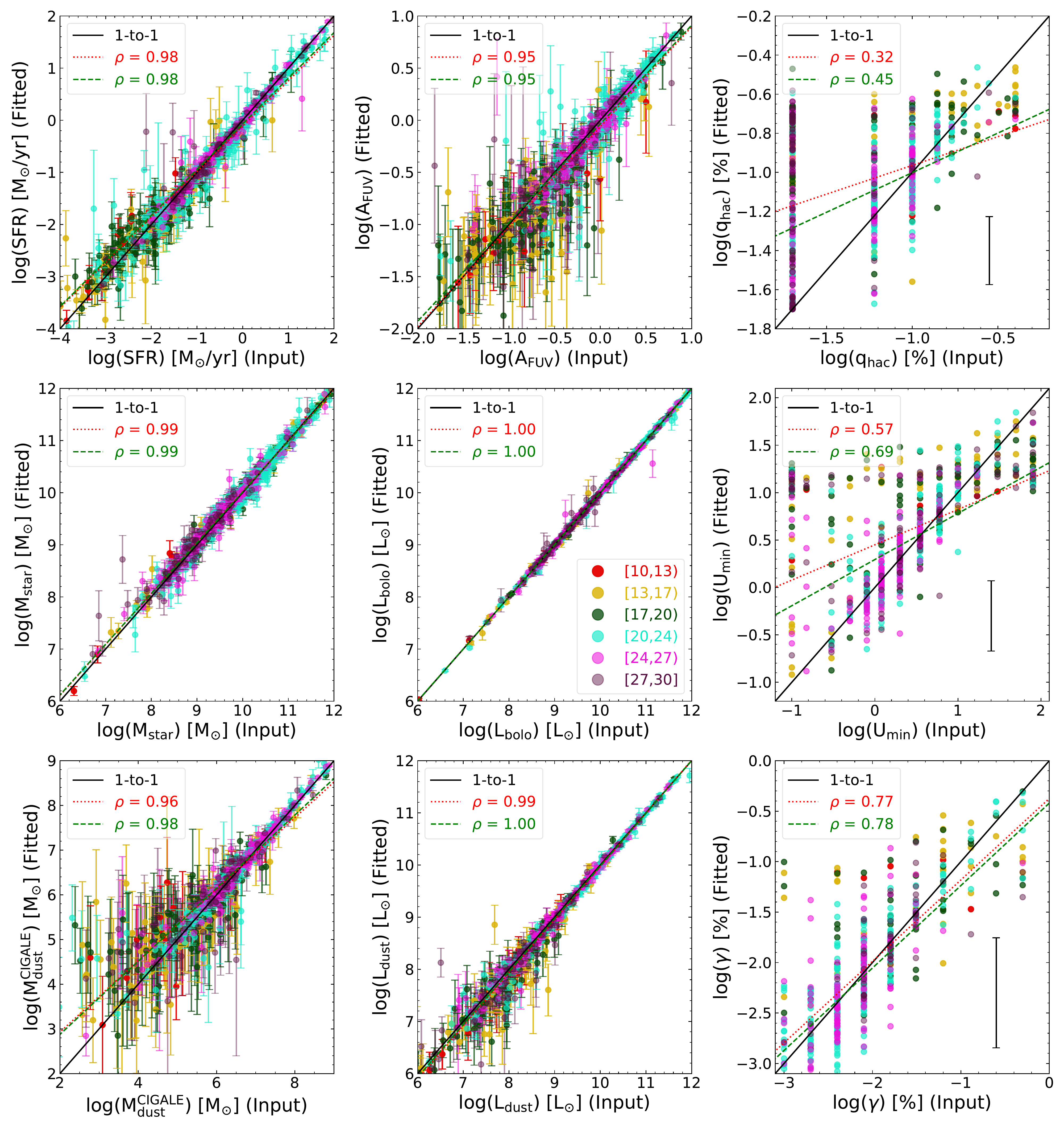}
\caption{Comparison between the best fitted parameters (input values; \textit{x}-axis) and the mock parameters (fitted values; \textit{y}-axis) estimated from the \textsc{CIGALE} fitted PDFs for each parameter (SFR, $A_\mathrm{FUV}$, $q_\mathrm{hac}$, $M_\mathrm{star}$, $L_\mathrm{bolo}$, $U_\mathrm{min}$, $M_\mathrm{dust}^\textsc{CIGALE}$, $L_\mathrm{dust}$, and $\gamma$, from top to bottom). For each galaxy the probability-weighted mean value along with the associated standard-deviation of the mock values (error bars) are plotted on the \textit{y}-axis. In the cases of $q_\mathrm{hac}$, $U_\mathrm{min}$, and $\gamma$, the average standard-deviation is plotted in the bottom-right corner of each panel, in order to avoid confusion. The data are colour-coded with the number of observations available for each galaxy (see the inset of the panel in the middle for the explanation of the colours). The solid black line is the one-to-one relation, the red dotted line is the linear regression to the full set of data, while the green dashed line is the linear regression to the galaxies with more than 20 observations available. The Spearman's coefficient ($\rho$) of the linear regression fits is also provided in each panel.}
\label{fig:mock}
\end{figure}

In this section we present the results of the mock analysis. The results of this procedure are shown in Fig.~\ref{fig:mock} with the input values of each parameter plotted on the \textit{x}-axis and the probability-weighted mean value along with the associated standard-deviation of the fitted values (error bars) on the \textit{y}-axis. The data are colour-coded with the number of fluxes available for each galaxy (see the inset of the panel in the middle for the explanation of the colours). There are 17, 86, 134, 337, 152, and 88 galaxies with their available number of observations between [10,13), [13,17), [17,20), [20,24), [24,27), and [27,30] respectively. 

In general, most of the parameters are well correlated with the input parameters but there is significant scatter in the cases of $M_\mathrm{dust}^\textsc{CIGALE}$, $A_\mathrm{FUV}$, $q_\mathrm{hac}$, $U_\mathrm{min}$ and $\gamma$. Galaxies with large deviations from the one-to-one relation usually have red, gold or green colours indicating that the lack of observations is an important cause of this discrepancy. In most cases the linear fit is very close to the one-to-one relation indicating the ability of \textsc{CIGALE} to retrieve the input values. The exceptions are the values of $q_\mathrm{hac}$, and $U_\mathrm{min}$ with their linear regression fits deviating significantly from the one-to-one relation. This is also shown with the Spearman's coefficient, $\rho=0.32$ and $\rho=0.57$ respectively. The linear regression improves when only galaxies with a sufficient number of observations (more than 20) are considered (green dashed lines) with the associated Spearman's coefficients being slightly higher ($\rho=0.45$ and $\rho=0.69$ for $q_\mathrm{hac}$, and $U_\mathrm{min}$ respectively).

To understand the nature of these deviations, we performed a series of tests. First, a visual inspection of the fitted SEDs of the galaxies that are the most deviant from the one-to-one relation in the mock analysis revealed that most of these objects lack crucial photometric data near the peak of the dust emission (just left or right from the peak), making it quite hard to constrain the dust parameters. Moreover, many of these deviant galaxies have large uncertainties on their FIR-submm flux measurements which also adds up to the poor constraint of $U_\mathrm{min}$.

Finally, our \textsc{CIGALE} set up, with the specific parameter grid used for this study, was also used by Tr\v{c}ka et al. (2019, in preparation) to fit synthetic SEDs (from the FUV to the submm wavelengths) of galaxies from the EAGLE \citep[Evolution and Assembly of GaLaxies and their Environments;][]{2015MNRAS.446..521S} simulations. In this sample a complete set of measurements of 29 bands were available for \textsc{CIGALE} to fit resulting in a very accurate recovery of the input values for each parameter (Tr\v{c}ka et al., 2019 in preparation). From the above tests it becomes evident that the most important cause of the large uncertainties observed in some of the derived parameters is the poor wavelength coverage for these galaxies.

\section{\textsc{CIGALE} modelling using the DL14 dust grain model and comparison with the \textsc{THEMIS} model} \label{ap:dl14}

\begin{figure}[t]
\centering
\includegraphics[width=9cm]{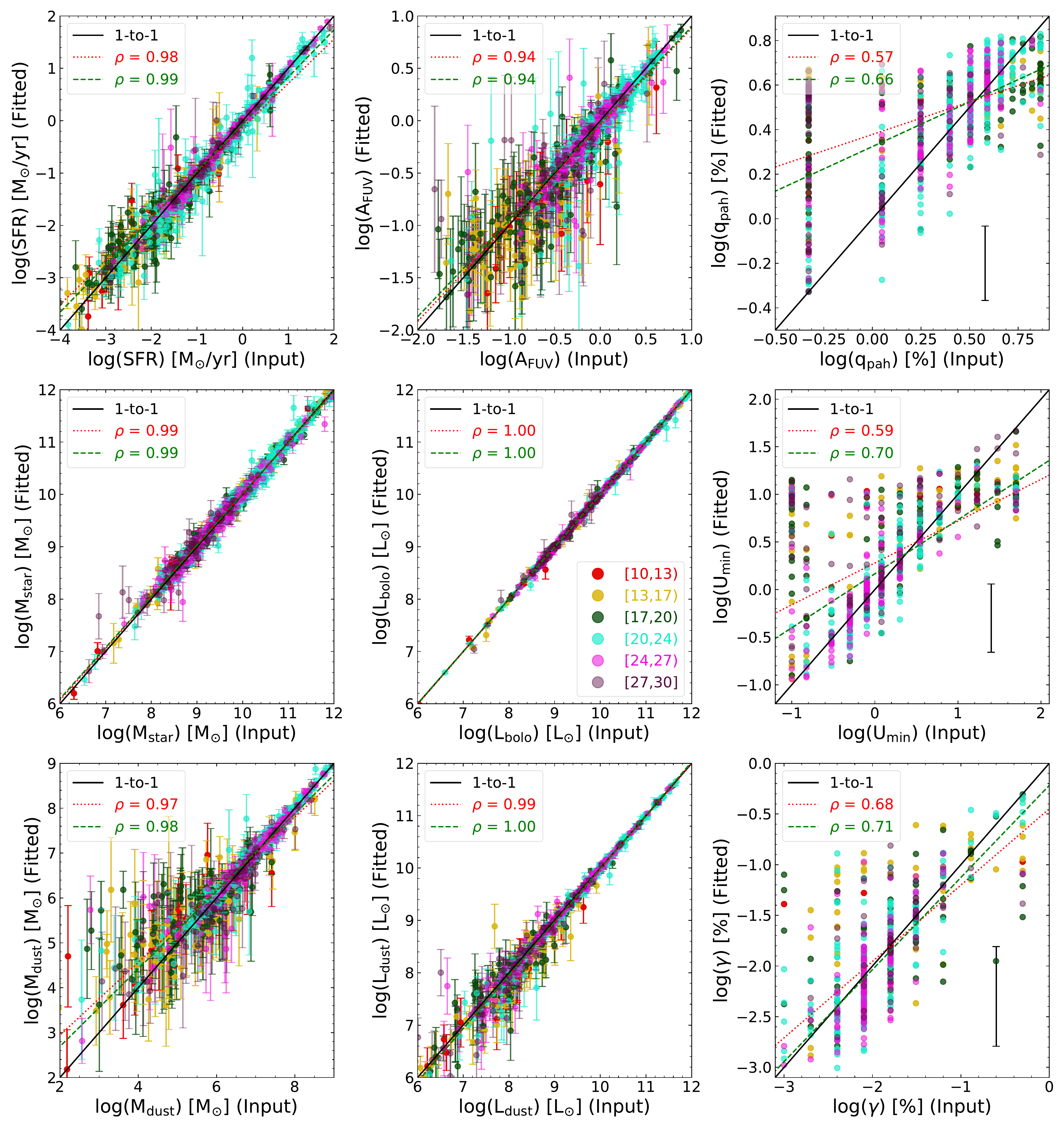}
\caption{Same as Fig.\ref{fig:mock} but for the DL14 dut model. In DL14, the PAH abundance ($q_\mathrm{PAH}$) is substituting the fraction of hydrocarbon solids ($q_\mathrm{hac}$).}
\label{fig:mock_dl14}
\end{figure}

Since \citet{2007ApJ...663..866D} \citep[updated in][DL14]{2014ApJ...780..172D} is a widely adopted model that describes the dust properties, we performed an additional fitting run with \textsc{CIGALE} using this model, instead of \textsc{THEMIS} (Sect.~\ref{subsubsec:par_space}). The parameter grid is the same as in the case of \textsc{THEMIS}, with only the PAH abundance ($q_\mathrm{PAH}$) substituting the fraction hydrocarbon solids ($q_\mathrm{hac}$; see Table~\ref{tab:param}). 

In Fig.~\ref{fig:mock_dl14} we show the results of the relevant mock analysis obtained from \textsc{CIGALE} with the input values of each parameter plotted on the \textit{x}-axis and the mock fitted values on the \textit{y}-axis. As in the case of Fig.~\ref{fig:mock}, data are colour-coded according to the number of observations available for each galaxy (see the inset of the panel in the middle for the explanation of the colours). As in the case where the \textsc{THEMIS} model is used, most of the input values of the parameters are well correlated with the values derived from the mock analysis with the exception of $q_\mathrm{PAH}$, and $U_\mathrm{min}$ that show a similar scatter. The Spearman's coefficients for these parameters are $\rho=0.57$ and $\rho=0.59$ respectively. The linear regression improves when only galaxies with a sufficient number of observations (more than 20) are considered (green dashed lines) with the associated Spearman's coefficients being slightly higher ($\rho=0.66$ and $\rho=0.70$ for $q_\mathrm{pah}$ and $U_\mathrm{min}$ respectively).

\begin{figure}[t]
\includegraphics[width=9cm]{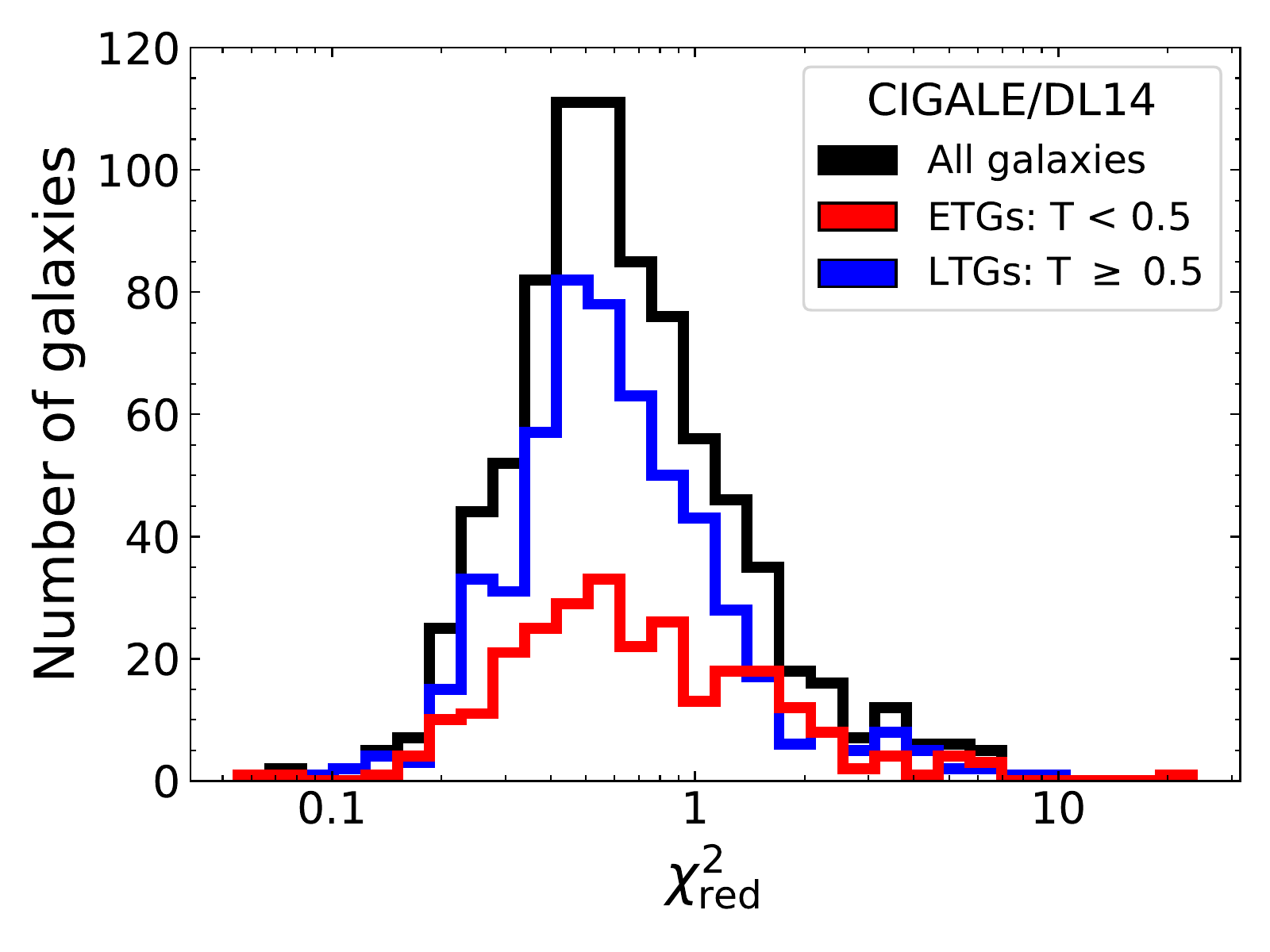}
\caption{Distribution of the reduced $\chi^2$ of the 814 galaxies modelled with \textsc{CIGALE} and with the DL14 dust model (black line). The distributions for the LTGs and the ETGs are also shown (blue and red lines respectively).}
\label{fig:chisqr_dl14}
\end{figure}

As for the \textsc{THEMIS} model we provide here the $\chi^2_\mathrm{red}$ distribution in Fig.~\ref{fig:chisqr_dl14} (to be compared with Fig.~\ref{fig:chisqr}). In the DL14 case the median value of $\chi^2_\mathrm{red}$ is 0.58 for the full sample, with 0.56 for the LTGs only and 0.61 for the ETGs. There are 57 ($\sim7\%$) galaxies with $\chi^2_\mathrm{red}>2$ and 19 ($\sim2\%$) with $\chi^2_\mathrm{red}>4$. 

\begin{figure}[t]
\centering
\includegraphics[width=9cm]{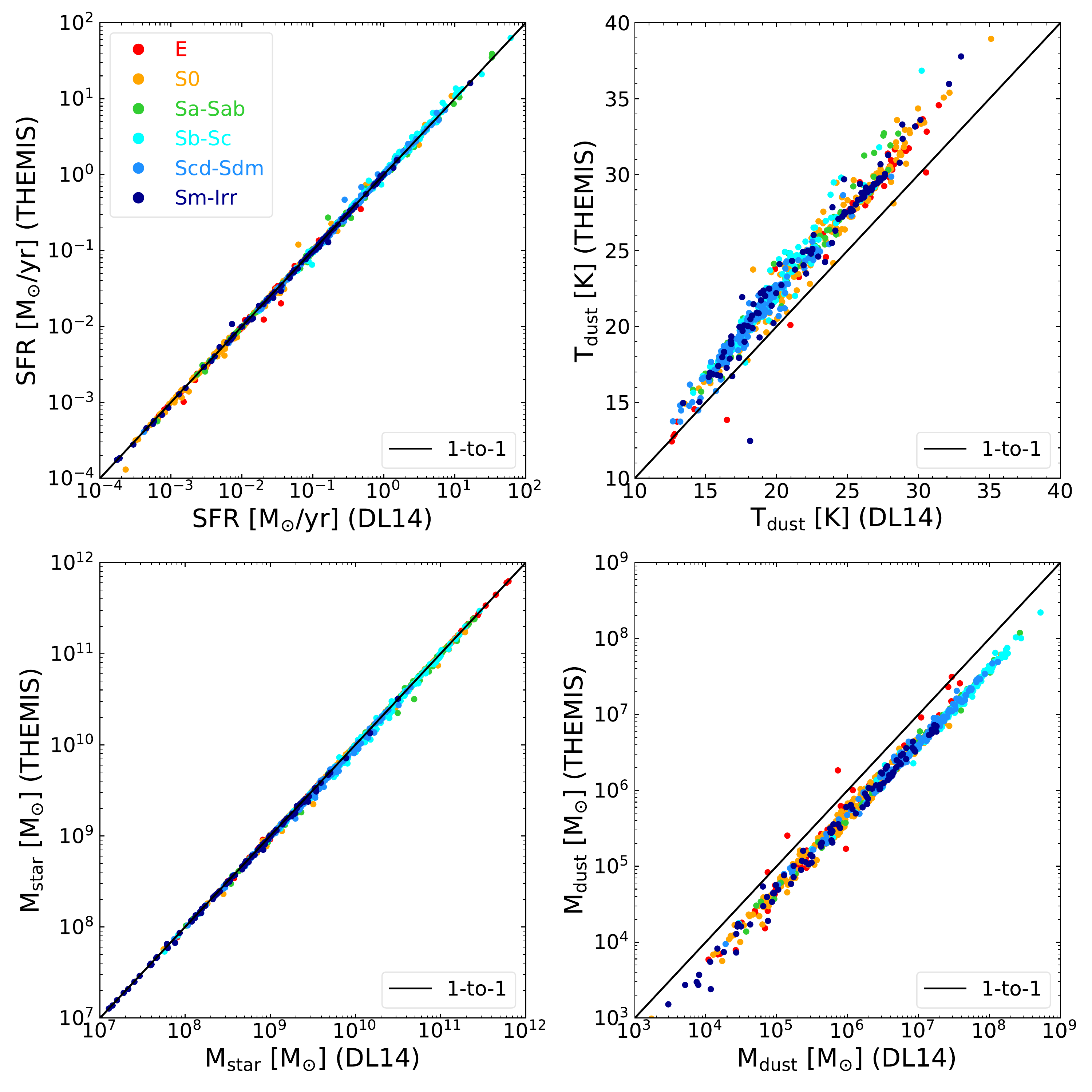}
\caption{Comparison between SFR, $T_\mathrm{dust}$, $M_\mathrm{star}$, and $M_\mathrm{dust}$ (from top-left to bottom-right respectively), as derived from \textsc{CIGALE}, assuming two different dust grain models (DL14 on \textit{x}-axis and \textsc{THEMIS} on \textit{y}-axis). The points are colour-coded according to the six main morphological types (see the inset in the top-left panel). The black solid line is the one-to-one relation.}
\label{fig:comp_dl14}
\end{figure}

In Fig.~\ref{fig:comp_dl14} we compare the results obtained by fitting \textsc{CIGALE} with the DL14 dust model (\textit{x}-axis) and the \textsc{THEMIS} model (\textit{y}-axis) for four parameters (SFR, $T_\mathrm{dust}$, $M_\mathrm{star}$, and $M_\mathrm{dust}$, from top to bottom respectively). In each panel the galaxies modelled by \textsc{CIGALE} are plotted, colour-coded with their morphology (see the inset in the top-left panel). It immediately becomes evident that parameters that are constrained, mainly, from the optical part of the SED of the galaxy (SFR and $M_\mathrm{star}$) are in very good agreement, and on the one-to-one relation, almost unaffected by the choice of the dust model. The dust parameters on the other hand, $T_\mathrm{dust}$ and $M_\mathrm{dust}$, depend on the choice of the dust model used and this is revealed by an offset from the one-to-one relation for all morphologies. In particular, DL14 predicts higher dust masses with a percentage difference of $\sim42\%$ and lower dust temperatures of $\sim5\%$. This offset is due to the fact that \textsc{THEMIS} is more emissive than DL14. \textsc{THEMIS} has both a lower $\beta$ and a higher $\kappa_0$ value \citep[e.g., Fig.~4 of][]{2018ARA&A..56..673G}.

\section{Comparison with different recipes used in the literature}\label{ap:comp}

\begin{figure}[t]
\centering
\includegraphics[width=9cm]{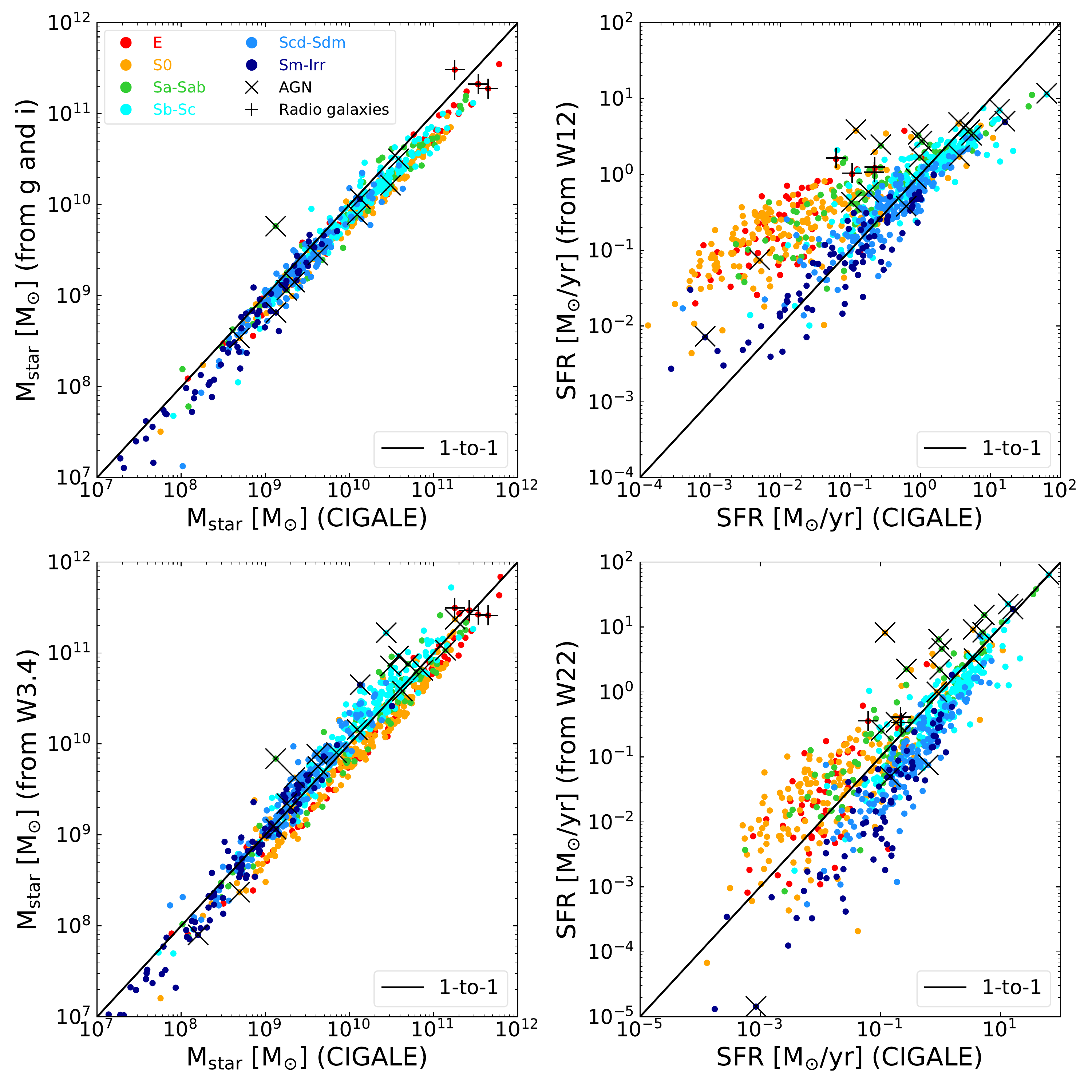}
\caption{Comparison between $M_\mathrm{star}$ (left panels) and SFR (right panels) derived from \textsc{CIGALE} (\textit{x}-axis), and other recipes widely used in the literature (\textit{y}-axis). The formulas used in the literature are given in the text. The points are colour-coded according to the morphological type, while galaxies hosting AGNs and strong radio-jets are marked with an `X' and a `+' respectively. The black solid line is the one-to-one relation.}
\label{fig:comp_other}
\end{figure}

Here we investigate how the parameters derived from \textsc{CIGALE} compare with values obtained from recipes, widely used in the literature. To obtain an alternative estimate of stellar mass we have used the formula derived in \citet{2013MNRAS.433.2946W}:

\begin{equation} \label{eq:mstar_wen}
\begin{gathered}
\log\left(\frac{M_\mathrm{star}}{\mathrm{M}_{\odot}}\right) = \left(-0.040 \pm 0.001\right) + \left(1.120 \pm 0.001\right) \\
\ \ \ \ \ \ \ \times \ \log\left(\frac{\nu L_\mathrm{\nu}(3.4~\mu\mathrm{m})}{\mathrm{L}_{\odot}}\right) \, ,
\end{gathered}
\end{equation}

\noindent where $L_\mathrm{\nu}(3.4~\mu\mathrm{m})$ is the WISE $3.4~\mu$m luminosity in L$_{\odot}$. Another recipe we used to determine the stellar mass was the formula derived in \citet{2011MNRAS.418.1587T}:

\begin{equation} \label{eq:mstar_taylor}
\begin{gathered}
\log\left(\frac{M_\mathrm{star}}{\mathrm{M}_{\odot}}\right) = 1.152 + 0.7\left(g-i\right) - 0.4M_i \, ,
\end{gathered}
\end{equation}

\noindent where \textit{g} and \textit{i} are the apparent magnitudes derived from our photometry of SDSS data in the respective bands and $M_i$ is the absolute $i$-band magnitude. Stellar masses derived from the above are compared with those obtained from \textsc{CIGALE} in Fig.~\ref{fig:comp_other} (left panels). In all panels of Fig.~\ref{fig:comp_other}, the circles are individual galaxies colour-coded with their morphology, as shown in the inset in the top-left panel, while the `X' and `+' symbols indicate AGN and strong radio galaxies (in many cases being extreme outliers in the correlations under investigation). Even though \citet{2011MNRAS.418.1587T} assumed a \citet{2003PASP..115..763C} IMF to retrieve the stellar masses, the comparison with \textsc{CIGALE} is fairly good but with a small offset. On the other hand, the stellar masses derived from the WISE $3.4~\mu$m band compare even better with the stellar masses derived from \textsc{CIGALE}. Thus, we can confirm that the recipe of \citet{2013MNRAS.433.2946W} is a good proxy for the stellar mass of a galaxy. 

We have used the WISE $12~\mu$m and WISE $22~\mu$m data to make alternative estimates of the SFR of the galaxies. Before calculating the SFR we have subtracted an estimate of the stellar continuum emission in the $12~\mu$m and $22~\mu$m bands using the data in Table~B1 of \citet{2014A&A...565A.128C}. They calculate the contamination separately for early and late type galaxies based on fits to the stellar continuum at shorter wavelengths. We used the values normalized to the IRAC $3.6~\mu$m band. \citet{2014ApJ...782...90C} have used the WISE $12~\mu$m data, calibrated against H${\alpha}$, to derive SFRs for GAMA sample galaxies matched to the WISE data \citep{2011MNRAS.413..971D}. These galaxies have a wide range of morphological types, though of course the initial calibration does require a measured H${\alpha}$ flux. Typically, the strongest individual contributor to the WISE $12~\mu$m pass band is the $11.3~\mu$m PAH feature, which is predominantly excited by ultraviolet radiation from young stars and hence the link to the SFR \citep{2008ApJ...684..270K}. \citet{2007ApJ...671..323H} and \citet{2007ApJ...667..149F} have previously shown that PAH features can be used as indicators of the current SFR. To this end we use the re-calibration of the \citet{2014ApJ...782...90C} WISE $12~\mu$m flux density SFR relation given in \citet{2016MNRAS.461..458D}:

\begin{equation} \label{eq:sfr_davies}
\begin{gathered}
\log\left(\frac{\text{SFR}}{\mathrm{M}_{\odot}\mathrm{yr}^{-1}}\right) = \left(0.66 \pm 0.01\right) \left[\log\left(L_{12}\right) - 22.25\right] \\
\ \ \ \ \ \ \ + \ \left(0.160 \pm 0.004\right) \, ,
\end{gathered}
\end{equation}

\noindent where $L_{12}$ is the WISE $12~\mu$m flux density in W~Hz$^{-1}$. This calibration has been updated onto a common standard for various SFR indicators. We note that the re-calibration is based on the properties of `typical' spiral galaxies (disc galaxies of stellar mass $9<\log\left(M_\mathrm{star}/\mathrm{M}_{\odot}\right)< 11$), though the original data and calibration includes galaxies with a larger range of morphological types. 

\citet{2015ApJS..219....8C} also provided calibrations for the star-formation by analyzing the MIR properties of the full SDSS spectroscopic galaxy sample. Of that work we used the WISE $22~\mu$m band SFR proxy:

\begin{equation} \label{eq:sfr_chang}
\begin{gathered}
\log\left(\frac{\text{SFR}}{\mathrm{M}_{\odot}\mathrm{yr}^{-1}}\right) = \log\left(L_{22}/\mathrm{L}_{\odot}\right) - 9.08 \, ,
\end{gathered}
\end{equation}

\noindent The comparison of calculated SFRs is shown in Fig.~\ref{fig:comp_other} (right panels). The fit is clearly good for the later types, but not so good for the early type galaxies. The $12~\mu$m flux calibration clearly overestimates the SFR of the ETGs compared to that obtained from \textsc{CIGALE}. Compared to literature values, i.e. \citet{2014MNRAS.444.3427D}, the $12~\mu$m values for ETGs are high, thus the \textsc{CIGALE} values are consistent with previous measures and so we accept and use these. On the other hand, the $22~\mu$m calibration underestimates the SFR of the latest type of galaxies while it also overestimates the SFR of the early types, however not to the same extent as for the $12~\mu$m. In both cases, it is apparent that MIR monochromatic band proxies, despite that they trace the warm dust and consequentially SFR, are not sufficient to get a good estimate of the current SFR.

\section{Recipes to estimate physical properties of galaxies as a function of Hubble stage ($T$)} \label{ap:spline}

\begin{table*}
\caption{Recipes to estimate the integrated physical properties of galaxies given their Hubble stage ($T$).}
\begin{center}
\scalebox{0.8}{
\begin{tabular}{llcccccc}
\hline 
\hline 
\multicolumn{8}{c}{$y = \alpha_0 + \alpha_1 \times T + \alpha_2 \times T^2 + \alpha_3 \times T^3 + \alpha_4 \times T^4 + \alpha_5 \times T^5$}\\
\hline
 $y$ & & $\alpha_0$ & $\alpha_1$ & $\alpha_2$ & $\alpha_3$ & $\alpha_4$ & $\alpha_5$ \\
\hline
$\log\left(M_\mathrm{dust}/M_\mathrm{star}\right)$ &               & $-4.01 \pm 0.07$ & $0.25 \pm 0.03$ & $0.01 \pm 0.01$ & $(-2.06 \pm 1.20)\times10^{-3}$ & $(-2.52 \pm 3.70)\times10^{-4}$ & $(1.93 \pm 2.85)\times10^{-5}$ \\
$\log\left(M_\mathrm{HI}/M_\mathrm{star}\right)$   &               & $-1.86 \pm 0.10$ & $0.18 \pm 0.04$ & $0.03 \pm 0.02$ & $(-0.68 \pm 1.84)\times10^{-3}$ & $(-6.30 \pm 5.80)\times10^{-4}$ & $(4.15 \pm 4.40)\times10^{-5}$ \\
$\log\left(\text{SFR}/M_\mathrm{star}\right)$      & [yr$^{-1}$]   & $-11.65\pm 0.06$ & $0.38 \pm 0.02$ & $0.03 \pm 0.01$ & $(-6.89 \pm 1.12)\times10^{-3}$ & $(-2.50 \pm 3.50)\times10^{-4}$ & $(4.66 \pm 2.70)\times10^{-5}$ \\
$\log\left(M_\mathrm{dust}/M_\mathrm{HI}\right)$   &               & $-2.25 \pm 0.10$ & $0.08 \pm 0.04$ & $-0.01\pm 0.02$ & $(-1.07 \pm 1.77)\times10^{-3}$ & $(-0.10 \pm 5.60)\times10^{-4}$ & $(0.79 \pm 4.30)\times10^{-5}$ \\
$\log\left(M_\mathrm{star}\right)$                 & [M$_{\odot}$] & $10.31 \pm 0.10$ & $0.07 \pm 0.04$ & $-0.01\pm 0.02$ & $(-4.71 \pm 1.90)\times10^{-3}$ & $(5.30  \pm 6.00)\times10^{-4}$ & $(-1.65\pm 4.60)\times10^{-5}$ \\
$\log\left(M_\mathrm{dust}\right)$                 & [M$_{\odot}$] & $6.28  \pm 0.12$ & $0.33 \pm 0.05$ & $-0.02\pm 0.02$ & $(-5.96 \pm 2.20)\times10^{-3}$ & $(6.36  \pm 6.80)\times10^{-4}$ & $(-3.03\pm 5.20)\times10^{-5}$ \\
$\log\left(M_\mathrm{HI}\right)$                   & [M$_{\odot}$] & $8.44  \pm 0.15$ & $0.18 \pm 0.06$ & $0.04 \pm 0.02$ & $(-4.77 \pm 2.60)\times10^{-3}$ & $(6.26  \pm 8.14)\times10^{-4}$ & $(5.70 \pm 6.20)\times10^{-5}$ \\
$\log\left(\text{SFR}\right)$                   & [M$_{\odot}$/yr] & $-1.33 \pm 0.10$ & $0.42 \pm 0.05$ & $0.02 \pm 0.02$ & $(-10.10\pm 2.00)\times10^{-3}$ & $(-1.25 \pm 6.20)\times10^{-4}$ & $(5.40 \pm 4.80)\times10^{-5}$ \\
$T_\mathrm{dust}^\textsc{CIGALE}$                  & [K]           & $24.47 \pm 0.52$ & $-1.01\pm 0.21$ & $0.10 \pm 0.08$ & $(0.66  \pm 9.12)\times10^{-3}$ & $(-2.84 \pm 2.87)\times10^{-3}$ & $(3.14 \pm 2.20)\times10^{-4}$ \\
 & & & & & & & \\
$f_\mathrm{old}^\mathrm{unatt}$   & & $0.98 \pm 0.01$ & $-0.030 \pm 0.004$ & $-0.010 \pm 0.002$ & $(5.22  \pm 1.70)\times10^{-4}$ & $(1.74  \pm 0.54)\times10^{-4}$ & $(-1.43 \pm 0.41)\times10^{-5}$ \\
$f_\mathrm{young}^\mathrm{unatt}$ & & $0.02 \pm 0.01$ & $ 0.030 \pm 0.004$ & $0.010  \pm 0.002$ & $(-5.22 \pm 1.70)\times10^{-4}$ & $(-1.74 \pm 0.54)\times10^{-4}$ & $(1.43  \pm 0.41)\times10^{-5}$ \\
$f_\mathrm{old}^\mathrm{att}$     & & $0.88 \pm 0.02$ & $-0.063 \pm 0.006$ & $-0.007 \pm 0.003$ & $(1.45  \pm 0.28)\times10^{-3}$ & $(1.29  \pm 0.88)\times10^{-4}$ & $(-1.69 \pm 0.68)\times10^{-5}$ \\
$f_\mathrm{young}^\mathrm{att}$   & & $0.01 \pm 0.02$ & $ 0.004 \pm 0.002$ & $0.001  \pm 0.001$ & $(0.07  \pm 0.10)\times10^{-3}$ & $(-0.04 \pm 0.31)\times10^{-4}$ & $(-0.01 \pm 0.24)\times10^{-5}$ \\
$f_\mathrm{abs}$                  & & $0.10 \pm 0.02$ & $ 0.056 \pm 0.007$ & $0.005  \pm 0.003$ & $(-1.43 \pm 0.31)\times10^{-3}$ & $(-1.18 \pm 0.97)\times10^{-4}$ & $(1.51  \pm 0.74)\times10^{-5}$ \\
\hline \hline
\end{tabular}}
\label{tab:spline}
\end{center}
\end{table*}

As already discussed in the main text, we have interpolated the median values, per morphological bin, of several physical parameters with a 5th order polynomial (see Figs.~\ref{fig:dsr} and \ref{fig:old_young_dust}). In Table~\ref{tab:spline} we provide the exact values of the coefficients of the polynomial regression. This allows us, for a galaxy of a given Hubble stage ($T$), to estimate the values of $M_\mathrm{star}$, $M_\mathrm{dust}$, $M_\mathrm{HI}$, SFR, $T_\mathrm{dust}^\textsc{CIGALE}$, the fractions of the stellar populations $f_\mathrm{old}^\mathrm{unatt}$, $f_\mathrm{young}^\mathrm{unatt}$, as well as the fractions of the attenuated luminosities of the old and the young stellar components $f_\mathrm{old}^\mathrm{att}$, $f_\mathrm{young}^\mathrm{att}$ and the fraction of the absorbed, by the dust, luminosity $f_\mathrm{abs}$. The typical uncertainty values of each polynomial coefficient was derived through bootstrapping the data. More specifically, for every relation we created 100 new data-sets by varying randomly the original values within their typical uncertainties. For every new data-set we computed the median values per morphological bin and fitted a 5th order polynomial through them. Then, by calculating the standard deviation of these 100 fitted lines we were able to get a measurement of the typical errors. The values are also provided in Table\ref{tab:spline}.


\end{document}